# Handling multivariable missing data in causal mediation analysis estimating interventional effects

S. Ghazaleh Dashti[1,2], Katherine J. Lee[1,2], Julie A. Simpson[3,4], John B. Carlin[1,2,3], Margarita Moreno-Betancur[1,2]

1 Clinical Epidemiology and Biostatistics Unit, Murdoch Children's Research Institute, Melbourne, Victoria, Australia
2 Department of Pediatrics, University of Melbourne, Melbourne, Victoria, Australia
3 Centre for Epidemiology and Biostatistics, Melbourne School of Population and Global Health, University of Melbourne, Melbourne, Victoria, Australia
4 Nuffield Department of Medicine, University of Oxford, Oxford, UK.

Corresponding author
S Ghazaleh Dashti, DDS, MPH, PhD
Clinical Epidemiology and Biostatistics Unit
Murdoch Children's Research Institute
Royal Children's Hospital, 50 Flemington Rd, Parkville VIC 3052
Phone: +61 38341 6200
Email: ghazaleh.dashti@mcri.edu.au

## Abstract

The interventional effects approach to causal mediation analysis is increasingly common in epidemiologic research given its potential to address policy-relevant questions about hypothetical mediator interventions. Multiple imputation is widely used for handling multivariable missing data in epidemiologic studies. However, guidance is lacking on best practice for using multiple imputation when estimating interventional mediation effects, specifically regarding the role of missingness mechanism in the performance of the method, how to appropriately specify the multiple imputation model when g-computation is used for effect estimation, and appropriate variance estimation. To address this gap, we conducted simulations based on the Victorian Adolescent Health Cohort Study. We considered seven missingness mechanisms, involving varying assumptions regarding the influence of an intermediate confounder, a mediator, and/or the outcome on missingness in key variables. We compared the performance of complete-case analysis, six multiple imputation approaches by fully conditional specification, differing in how the imputation model was tailored, and a "substantive model compatible" multiple imputation-fully conditional specification approach. We evaluated MIBoot (multiple imputation, then bootstrap) and BootMI (bootstrap, then multiple imputation) approaches for variance estimation. All multiple imputation approaches, apart from those clearly diverging from best practice, yielded approximately unbiased estimates when none of the intermediate confounder, mediator, and outcome variables influenced missingness in any of these variables and non-negligible bias otherwise. We observed the largest bias for interventional effects when each of the intermediate confounders, mediators, and outcomes influenced their own missingness. BootMI returned variance estimates with smaller bias than MIBoot.

## Introduction

Broadly speaking, mediation analysis aims to investigate the role of intermediate factors on the path from an exposure to an outcome. Modern causal mediation analysis provided a formal approach to this problem rooted in the potential outcomes framework, making the critical distinction between estimand, identification, and estimation. The so-called natural indirect and direct effects were initially proposed as definitions for mediation estimands, but in recent years they have been criticized for their lack of translational potential. These effects require a "cross-world independence" identifiability assumption that cannot be verified or mapped to a hypothetical randomized experiment, and which renders them unidentifiable in settings with intermediate (exposure-induced) mediator–outcome confounding and, more generally, in settings with multiple mediators.(1)

Interventional effects have been introduced as alternative estimands that overcome these limitations.(2, 3) The interventional indirect effect seeks to quantify the extent to which the expected outcome under exposure may be altered by a hypothetical intervention shifting the distribution of the mediator(s) of interest to the corresponding distribution under no exposure. The interventional direct effect is defined as the remaining effect of exposure on outcome.(2) The interventional effects approach to causal mediation analysis is increasingly used in epidemiology to address policy-relevant questions concerning mediators that could potentially be targeted by interventions to reduce disease risk in the exposed, and thus reduce exposure-group disparities (e.g.,(4-7)). Under some identifiability assumptions and in the absence of missing data, g-methods can be used to estimate the interventional mediation effects.(2)

With missing data, additional identifiability assumptions and adaptation of estimation methods are required to address potential bias and loss of precision,(8) but research is scant on how to handle this problem when estimating interventional effects. For natural mediation effects, which are equivalent to interventional effects in the single-mediator setting with no intermediate confounding,(3) previous research has derived identification results and proposed estimation approaches within the instrumental variable framework when only the outcome variable has missing data.(9, 10) However, in epidemiologic studies, it is common to encounter multivariable missing data (missingness in the exposure, mediator, outcome, and confounders). In such cases, whether estimating natural or interventional effects, principled and more flexible approaches for handling multivariable missing data are necessary.

Multiple imputation is one such approach. It involves creating multiple complete datasets, by replacing missing values with draws from approximate posterior distributions given the observed data. Generally, each imputed dataset is analyzed separately and the results are pooled to obtain the final multiple imputation estimate and standard error (SE).(11) An advantage of multiple imputation compared with performing analysis on complete records (i.e., complete-case analysis), is that it can incorporate auxiliary variables (predictors of missing values not in the substantive analysis) that may be helpful in gaining precision and/or reducing bias, depending on the missingness mechanism.(8)

In the context of natural effects, previous simulation studies have shown that multiple imputation and complete-case analysis return estimates with negligible bias in settings where the outcome and/or mediator is missing completely at random (MCAR, missingness not influenced any variable), while multiple imputation also has negligible bias when outcome and/or mediator missingness is only influenced by the fully observed exposure variable, which is a restricted version of missing at random (MAR), but not when these variables are missing not at random (MNAR, missingness influenced by the outcome/mediator variables).(10, 12-15) Aside from the focus on natural effects, none of these papers consider settings where the exposure and confounders also have missing data. Further, they focus on the restrictive "product method" for mediation analysis,(16) with no consideration of the modern g-methods which allow flexible model specification. This may explain the lack of research on how to ensure that the imputation model remains compatible with the mediation analysis, which is known in the multiple imputation literature to be essential to avoid bias. In general terms, compatibility requires the imputation model to incorporate all variables and assumed relationships in the substantive analysis, including interactions and non-linear terms.(8) The focus on the product method may also explain the lack of attention to the issue of variance estimation, with the default approach for obtaining point and variance estimates from multiple imputed datasets being to use Rubin's rules. However, the validity of Rubin's rules is contingent on the compatibility of the imputation model with the substantive analysis.(17) Also, the applicability of Rubin's rules with g-methods is unclear as they are not maximum likelihood estimators.(18)

Classifying missingness mechanisms using the MCAR/MAR/MNAR framework also has limitations. In multivariable missingness settings, understanding and evaluating the MAR assumption is not straightforward. Also, although MAR is sufficient, it is not necessary for unbiased estimation.(14, 19-21) For example, in the context of mediation analysis, Zuo and colleagues have demonstrated the recoverability (i.e., identifiability) of natural effects in the presence of missing data in the mediator and/or outcome for certain MNAR settings under further assumptions.(21) However, as noted by them, as well as other researchers in other contexts (10, 14, 20, 21) demonstrating recoverability would be complicated in more general settings, for example when there is also missingness in the confounders, and would often require imposing further assumptions, which in turn would limit the relevance of the recoverability results to very specific contexts. Further, in point exposure epidemiologic studies, previous simulation studies have shown that even under missingness mechanisms where the target estimand is shown to be non-recoverable, multiple imputation may still return approximately unbiased estimates.(14, 20, 22)

In the present paper, we seek to extend the previous research on methods for handling missing data in mediation analyses. Using a simulation study based on data from the Victorian Adolescent Health Cohort Study (VAHCS),(23) we aim to fill existing gaps by considering the estimation of interventional mediation effects in the context of one mediator of interest and one intermediate confounder, and missing data in exposure, outcome, mediator, and confounders. Our overarching aim is to provide guidance on considerations and best practice for using multiple imputation in this context, as it is a flexible, readily available, and widely used missing data method. Specifically, we make the following contributions. First, we evaluate the performance of multiple imputation when estimating interventional effects with intermediate confounding under seven general multivariable missingness mechanisms that are plausible in epidemiological studies, specified using missingness directed acyclic graphs (m-DAGs; DAGs with additional nodes representing missingness indicators for each variable with missing data).(20, 24) This alternative to the MCAR/MAR/MNAR framework facilitates the communication and assessment of multivariable missingness assumptions in practice. Second, we investigate how to appropriately specify the imputation model to avoid bias in estimation arising from incompatibility with the mediation analysis in the context of g-methods, specifically g-computation. Finally, we investigate the appropriate approach for variance estimation in mediation analysis when using multiple imputation, comparing the performance of Rubin's rules with an alternative approach proposed by von Hippel and Bartlett that can be expected to provide nominal coverage for non-maximum likelihood estimators and in the absence of compatibility.(25)

In the following sections, we introduce the case study and review methods for mediation analysis using interventional effects. We then present our simulation study and revisit the analysis of the VAHCS case study in light of our findings.

## Case study

The case study was based on a previous investigation using data from VAHCS, a longitudinal cohort study of 1,943 participants recruited at ages 14-15 years in 1992 and followed up into their adulthood across 10 waves of data collection (participants were 34-35 years old at wave 10). The study was approved by the Human Research Ethics Committee of the Royal Children's Hospital. The investigators were interested in the extent to which the impact of common mental disorders (CMD) in adolescence (waves 2-6; the exposure) on CMD mid-adulthood (wave 10, the outcome) could be countered by hypothetical interventions on potential mediators measured in young adulthood (waves 7-9).(23) Here we focused on CMD in young adulthood as the mediator of interest, which had a relatively large estimated interventional indirect effect, and on parenthood by age 24 as the intermediate confounder.(23)

In the analysis of the case study, we considered a broad set of pre-exposure confounders, and auxiliary variables as in the original investigation (see section Illustration with the case study). In the simulation study, the design of which was based on the case study, we considered a reduced set of pre-exposure confounders (adolescent cannabis use and anti-social behavior (waves 2-6)), and auxiliary variables (parent smoking by wave 7) for simplicity (Figure 1). As in the original investigation, for the case study, the analytic sample included 1,923 participants after excluding 20 individuals who died during the follow-up. The descriptive statistics in the sample for the variables used in the simulation study are shown in Table 1. Missingness proportions ranged from 0% for sex to 34% for cannabis use.

## Mediation analysis using interventional effects

Following the proposal in Moreno-Betancur et al,(2) we define the interventional indirect effect in the VAHCS example as the reduction in the risk of mid-adulthood CMD (outcome) among those with CMD in adolescence (exposed group), achievable by a hypothetical intervention that independently of parenthood by age 24 (intermediate confounder), would shift the distribution of young adulthood CMD (mediator) in those exposed (given pre-exposure confounders) to the levels among individuals without adolescent CMD (given pre-exposure confounders), while the distribution of parenthood by age 24 would remain as is in the exposed group (no intervention on this variable has been conceptualized). The interventional direct effect captures the remaining gap in risk of mid-adulthood CMD in the exposed compared with the unexposed group following the above-described intervention.

Without missing data, and under consistency, conditional independence, and positivity assumptions (see references (2, 26) for detailed explanation of these assumptions) these effects are identifiable by the following two expressions, where $X$, $L$, $Z$, and $Y$ denote binary exposure, intermediate confounder, mediator, and outcome variables respectively (1: present, 0: absent), and $\boldsymbol{C}$ a vector of measured confounders:

The interventional indirect effect:

$$\sum_{\boldsymbol{c}}\sum_{l}\sum_{z} \mathrm{P}(Y|X=1,Z=z,L=l,\boldsymbol{C}=\boldsymbol{c})\mathrm{P}(Z=z,L=l|\boldsymbol{C}=\boldsymbol{c})\mathrm{P}(\boldsymbol{C}=\boldsymbol{c}) - \sum_{\boldsymbol{c}}\sum_{l}\sum_{z} P(Y|X=1,Z=z,L=l,\boldsymbol{C}=\boldsymbol{c})\mathrm{P}(Z=z|X=0,\boldsymbol{C}=\boldsymbol{c})\,\mathrm{P}(L=l|X=1,\boldsymbol{C}=\boldsymbol{c})\mathrm{P}(\boldsymbol{C}=\boldsymbol{c}) \quad \text{(i)}$$

The interventional direct effect:

$$\sum_{\boldsymbol{c}}\sum_{l}\sum_{z} P(Y|X=1,Z=z,L=l,\boldsymbol{C}=\boldsymbol{c})\mathrm{P}(Z=z|X=0,\boldsymbol{C}=\boldsymbol{c})\,\mathrm{P}(L=l|X=1,\boldsymbol{C}=\boldsymbol{c})\mathrm{P}(\boldsymbol{C}=\boldsymbol{c}) - \sum_{\boldsymbol{c}}\sum_{l}\sum_{z} \mathrm{P}(Y|X=0,Z=z,L=l,\boldsymbol{C}=\boldsymbol{c})\mathrm{P}(Z=z,L=l|\boldsymbol{C}=\boldsymbol{c})\mathrm{P}(\boldsymbol{C}=\boldsymbol{c}) \quad \text{(ii)}$$

Estimation can proceed using a Monte Carlo (MC) (i.e. simulation-based) g-computation approach and standard errors (SEs) and 95% confidence intervals (CIs) can be obtained using bootstrap (see Box 1 for the steps involved).(3)

## Simulation Study

### *Aims*

1. To compare the bias of complete case analysis and multiple imputation approaches when estimating the interventional effects (i) and (ii) across a set of typical multivariable missingness scenarios depicted by m-DAGs (detailed below and in Figure 2).

2. To compare the performance different implementations of multiple imputation that differ in their compatibility with the g-computation approach used for estimating the interventional effects (detailed below and in Table 2).

3. To compare two approaches for combining multiple imputation and bootstrap for variance estimation: Rubin's rules (referred to as MIBoot here), and the approach proposed by von Hippel and Bartlett (referred to as BootMI). Details of implementing each approach are summarized in Box 2.

*Generating the complete data*

We conducted a simulation study based on the VAHCS case study. We simulated 2,000 complete datasets, each with 2,000 records and generated variables sequentially according to the DAG in Figure 1, based on the following models (all binary variables are coded 0/1 and $\text{logit}^{-1}((\cdot)) = \exp(\cdot)/(1+\exp(\cdot))$):

$A \sim \text{Bernoulli}\left(\text{logit}^{-1}(\alpha_0)\right)$

$C_1 \sim \text{Bernoulli}\left(\text{logit}^{-1}(\beta_0)\right)$

$C_2 \sim \text{Bernoulli}(\text{logit}^{-1}(\gamma_0 + \gamma_1 A))$

$C_3 \sim \text{Bernoulli}\left(\text{logit}^{-1}(\delta_0 + \delta_1 A)\right)$

$X \sim \text{Bernoulli}(\text{logit}^{-1}(\zeta_0 + \zeta_1 C_1 + \zeta_2 C_2 + \zeta_3 C_3 + \zeta_4 C_1 C_3))$

$L \sim \text{Bernoulli}(\text{logit}^{-1}(\eta_0 + \eta_1 X + \eta_2 C_1 + \eta_3 C_2 + \eta_4 C_3 + \eta_5 C_1 C_3))$

$Z \sim \text{Bernoulli}(\text{logit}^{-1}(\theta_0 + \theta_1 X + \theta_2 L + \theta_3 C_1 + \theta_4 C_2 + \theta_5 C_3 + \theta_6 C_1 C_3 + \theta_6 XL))$

$Y \sim \text{Bernoulli}(\text{logit}^{-1}(\omega_0 + \omega_1 X + \omega_2 L + \omega_3 Z + \omega_4 C_1 + \omega_5 C_2 + \omega_6 C_3 + \omega_7 XL + \omega_8 XZ + \omega_9 C_1 C_3))$

The parameter values in the models were determined by fitting similar models to the available VAHCS data (eTable 1).

*Imposing missing data*

We considered seven missingness scenarios depicted by the m-DAGs in Figure 2, where $M_{C_2}, M_{C_3}, M_X, M_L, M_Z, M_Y$ represent missingness indicators (1 if corresponding variable is missing, 0 otherwise) for $C_2, C_3, X, L, Z, Y$ respectively, and $W$ represents an unmeasured common cause for the missingness indicators. m-DAGs A to G depict missingness mechanisms that can be typically encountered in epidemiologic studies and for which we expected the performance of CCA and/or MI to be distinct based on previous research.(14, 20)

For each m-DAG, we imposed missingness on $C_2, C_3, X, L, Z, Y$ through generating the missingness indicators using the following models, where for m-DAGs in which an arrow from the variable to the missingness indicator was absent, we set the coefficients for the variable to 0, and in the presence of an arrow we set the coefficient to 0.9 (i.e., an odds ratio of 2.5, which represented a strong but potentially realistic association between variables and missingness indicators):

$W \sim \text{Bernoulli}\left(\text{logit}^{-1}(\rho_0)\right)$

$M_{C_2} \sim \text{Bernoulli}\left(\text{logit}^{-1}(\iota_0 + \iota_1 C_1 + \iota_2 C_2 + \iota_3 X + \iota_4 L + \iota_5 Z + \iota_6 Y + \iota_7 W)\right)$

$M_{C_3} \sim \text{Bernoulli}(\text{logit}^{-1}(\kappa_0 + \kappa_1 C_1 + \kappa_2 C_3 + \kappa_3 X + \kappa_4 L + \kappa_5 Z + \kappa_6 Y + \kappa_7 W))$

$M_X \sim \text{Bernoulli}\left(\text{logit}^{-1}(\lambda_0 + \lambda_1 C_1 + \lambda_2 C_2 + \lambda_3 C_3 + \lambda_4 X + \lambda_5 L + \lambda_6 Z + \lambda_7 Y + \lambda_8 W)\right)$

$M_L \sim \text{Bernoulli}(\text{logit}^{-1}(\mu_0 + \mu_1 C_1 + \mu_2 C_2 + \mu_3 C_3 + \mu_4 X + \mu_5 L + \mu_6 Z + \mu_7 Y + \mu_8 W))$

$M_Z \sim \text{Bernoulli}(\text{logit}^{-1}(\nu_0 + \nu_1 C_1 + \nu_2 C_2 + \nu_3 C_3 + \nu_4 X + \nu_5 L + \nu_6 Z + \nu_7 Y + \nu_8 W))$

$M_Y \sim \text{Bernoulli}(\text{logit}^{-1}(\xi_0 + \xi_1 C_1 + \xi_2 C_2 + \xi_3 C_3 + \xi_4 X + \xi_5 L + \xi_6 Z + \xi_7 Y + \xi_8 W))$

We modified intercepts and coefficient values for $W$ across models to keep the proportion with missing data, for each variable and overall, approximately the same across all scenarios and as in the VAHCS data (eTable 2).

*Analysis of the simulated data*

For each dataset in each missingness scenario, we applied Monte Carlo (MC) g-computation using correctly specified mediation and outcome models, in conjunction with the missing data approaches described below to estimate the interventional effects (i) and (ii). To reduce computational time, we performed MC g-computation using 50 Monte Carlo draws.

For handling missing data, we considered complete case analysis, six multiple imputation approaches within the fully conditional specification framework, and a substantive model compatible fully conditional specification multiple imputation approach, which is a modification of multiple imputation – fully conditional specification designed to facilitate imputing variables with missing data from models that are compatible with a specified substantive model.(27) In the mediation setting, there is no single “substantive model”, so how this approach should be implemented is not obvious. Here we evaluated an implementation that enforced compatibility with the outcome model (as opposed to the mediator or intermediate confounder model) used in g-computation. The approaches are summarized in Table 2 (see also eTable 3 for a list of variables and interaction terms included in the univariate imputation models for each multiple imputation – fully conditional specification approach). We used the R *mice* (28) and *smcfcs* (29) packages to perform multiple imputation. For each multiple imputation approach, we used MIBoot and BootMI (Box 2) to obtain the final multiple imputation point estimates and 95% confidence intervals. Following current recommendations, we generated 50 imputed datasets under the MIBoot and two under the BootMI approaches and drew 200 bootstrap samples with replacement for both. The R *bootImpute* package was used for the BootMI approach.(30) All analyses were performed in R version 4.2.2.(31) Simulation study codes can be downloaded from https://github.com/dashtisg/missingness_mediation.

### *Evaluation criteria*

For each missingness scenario, we generated and analyzed 2,000 datasets to estimate relative bias percent, empirical SE, percent error in average model-based SE relative to the empirical SE, and coverage probability for each analytic approach (combination of missing data approaches and approaches for variance estimation).(32) To obtain these metrics, the true values of the mediation effects were estimated by fitting correctly specified models to a simulated dataset of size 1,000,000 records with complete data, using MC g-computation with 1,000 Monte Carlo draws.

### *Simulation study results*

We estimated the true values of the indirect effect risk difference to be 6 per hundred and the direct effect risk differences to be 7 per 100.

Due to perfect prediction, the MI-SMCFCS (BootMI) approach failed in 13 out of the 2,000 simulated datasets under m-DAG B and in 7 out of 2,000 under m-DAG E.

#### Relative bias

For each multiple imputation approach, relative bias using BootMI and MIBoot were approximately the same. We present relative bias in point estimates for the indirect effects obtained using BootMI in Figure 3 and eFigure 1 (detailed results for both BootMI and MIBoot approaches are presented in eTable 4).

The relative bias of complete case analysis and multiple imputation approaches depended on the m-DAG. Within each m-DAG, the performance of the multiple imputation approaches did not vary greatly, except for MI-noLZY and MI-noY, which generally returned estimates with high bias across all m-DAGs (across both methods, relative bias ranging from within ±20% to ±81% for the indirect effect; within ±7% to ±38% for the direct effect).

For the indirect effect (Figure 3), complete case analysis had small bias under m-DAGs A and D (within ±6%) and large bias otherwise (-75% to -33%). Multiple imputation approaches, excluding MI-noLZY and MI-noY, had small bias under m-DAGs A, D, and F (within ±5%) and larger bias otherwise, but the bias was smaller under m-DAGs B and C (-23% to -11%) compared with m-DAGs E and G (-31 to -24%).

For the direct effect (eFigure 1), complete case analysis had small bias under m-DAGs A, B, and D (within ±13%) and high bias otherwise (-81% to -37%). MI approaches, excluding MI-noLZY and MI-noY, had small bias under m-DAGs A, B, D, F (0% to ±9%), moderate bias under m-DAG E (10% to 17%) and large bias under m-DAGs C and G (-36% to -19%).

Empirical standard error (SE), percent error in model-based SE, and coverage

For each missing data method, the empirical SEs were similar across the m-DAGs (eTable 4). For the indirect effect, the empirical SEs returned by complete case analysis (1.06 to 1.83 percentage points across the m-DAGs) were larger than for the multiple imputation approaches (0.44 to 1.62). Similarly, for the direct effect, complete case analysis returned larger SEs (2.93 to 3.47 percentage points) compared with the multiple imputation approaches (2.33 to 2.96). For each multiple imputation approach, MIBoot and BootMI returned comparable empirical SEs. Of the multiple imputation approaches, MI-noLZY and MI-noY performed similarly, yielding the smallest empirical SEs for both the indirect (0.45 to 1.05 percentage points) and direct effects (2.33 to 2.54 percentage points) and MI-SMCFCS yielded the largest empirical SEs (1.35 to 1.62 for the indirect effect and 2.66 to 2.99 percentage points for the direct effect).

For each method, the errors in estimated model SEs were remarkably similar across the m-DAGs (Figure 4 and eFigure 2). Complete class analysis returned model SEs with small error for both indirect (within ±5%) and direct effects (within ±2%). For the indirect effect (Figure 4), across all multiple imputation approaches, BootMI produced SEs with error within ±4%, while MIBoot produced SEs with larger error, with the largest error observed for MI-noLZY approach (ranging from 21% to 56%). For all other approaches, errors were within ±27%. For the direct effect also (eFigure 2), across all multiple imputation approaches, BootMI produced SEs within ±3% error, while MIBoot returned SEs with somewhat larger error (% error ranging from ±2% to ±19%).

As expected, the coverage probability of complete class analysis was close to nominal for both the indirect (eFigure 3) and direct (eFigure 4) effects where relative bias was small (m-DAGs A and D), and it was below nominal under other m-DAGs. Across all m-DAGs and all multiple imputation approaches, the coverage probability when using MIBoot was below nominal for the indirect and direct effects. When using BootMI with all multiple imputation approaches (except MI-noLZY and MI-noY), for the indirect effect the coverage probability was close to nominal in m-DAGs A, D, and F, and for the direct effect in m-DAGs A, B, D, E, F. For both effects, it was below nominal in other m-DAGs.

## Illustration with the case study

We illustrate how to apply the findings of the simulation study in the analysis of the case study. We first consider the missingness mechanism in the VAHCS example. If based on expert knowledge, we were able to make the assumption that the mediator(s) and outcome did not directly cause their own missingness (e.g., m-DAG D), then according to our simulation results all missing data approaches considered in the simulation study (except for MI-noLZY and MI-noY) may be expected to be approximately unbiased, with MI-SMCFCS (BootMI) being the preferred approach in terms of bias and variance for estimating the interventional indirect and direct effects using Monte Carlo g-computation. Results using all missing data approaches are shown in eFigure 5 for comparison. In all analyses, the Monte Carlo g-computation was implemented using 1,000 Monte Carlo draws. The BootMI approaches were conducted using 200 bootstrap samples and two imputed datasets, and the MIBoot approaches using 50 imputed datasets and 200 bootstrap samples. Following the original investigation,(23) sex, cannabis use, anti-social behavior, incomplete high school, parental completion of high-school, parental divorce, and socioeconomic disadvantage, all measured across adolescent waves (waves 2-6) were included as baseline confounders in the analysis. Parenthood by age 2 (wave 8) was included in the analysis as intermediate confounder. We included parent smoking and drinking by wave 7 and participants drinking and smoking at each of waves 2 to 6 as auxiliary variables in the multiple imputation approaches. There were missing data in some of the auxiliary variables (ranging from 5% to 21%) and additional baseline confounders (ranging from 0.6% to 7%). These variables were also multiply imputed in the all multiple imputation approaches.

Using MI-SMCFCS (BootMI), it was estimated that there were 14 per 100 additional cases of common mental disorders in mid-adulthood in those with vs without adolescent disorders (95%CI 8 to 21 per 100). A hypothetical intervention reducing the prevalence of young adult common mental disorders in individuals with adolescent disorders to that in individuals without adolescent disorders would result in 6 per 100 fewer cases of mid-adulthood disorders (95%CI 3 to 8 per 100). There would remain an additional 9 per 100 (95%CI 3 to 15 per 100) cases of mid-adulthood common mental disorders in individuals with vs without adolescent disorders following the intervention. The estimated effects were similar across the MI approaches except for MI-noZY, which returned a somewhat smaller estimate of the indirect effect. SEs estimated using BootMI were smaller than with MIBoot, but the difference between the two approaches was small.

If instead we considered that parenthood by age 24 ($L$), common mental disorders in young adulthood ($Z$) and mid-adulthood ($Y$) were likely to cause their own missingness, which is highly plausible in the VAHCS case study (e.g., m-DAG G), then none of these approaches to handling missing data can be expected to return approximately unbiased estimates of the indirect and direct effects. In settings such as this, researchers must interpret results from any of these approaches with extreme caution and consider conducting a sensitivity analysis such as delta-adjustment with multiple imputation using not-at-random fully conditional specification (NARFCS).(33, 34)

## Discussion

In this paper we have considered three issues to guide best practice when using multiple imputation to handle multivariable missing data in the context of causal mediation analysis estimating interventional [in]direct effects, specifically related to 1) the role of missingness mechanism in the performance of multiple imputation, 2) appropriate specification of the imputation model for compatibility with a causal mediation analysis using g-computation, and 3) the appropriate approach for variance estimation. First, we highlighted the important role of the missingness mechanism in the performance of missing data methods in terms of bias of the estimated effects. Ostensibly, best-practice multiple imputation approaches (i.e., MI-SMCFCS and appropriately tailored MI-FCS approaches) yielded approximately unbiased estimates when none of intermediate confounder, mediator and outcome variables influenced missingness in any of these variables, and non-negligible bias otherwise. For both interventional mediation effects, the highest bias was observed when each of the intermediate confounder, mediator and outcome caused their own missingness. Second, compatibility with the substantive analysis played a key role in ensuring that MI approaches performed well in terms of bias when estimating interventional mediation effects. To address this, researchers can use substantive model-compatible fully conditional specification multiple imputation enforcing compatibility with the outcome model in the g-computation approach, or fully conditional specification multiple imputation approaches incorporating all analysis variables (specifically the intermediate confounder, mediator and outcome variables in the mediation setting) in the imputation model. Third, the BootMI method was superior to MIBoot as it returned model SEs with smaller bias.

Our simulation results showed that for both interventional mediation effects, for missingness mechanisms where the intermediate confounder, mediator and outcome variables did not influence missingness in any of the intermediate confounder, mediator, or outcome variables (m-DAGs A, D, and F), standard best-practice MI approaches returned approximately unbiased estimates. For the indirect effect, these MI approaches were also approximately unbiased when intermediate confounder and mediator variables did not influence their own missingness (mDAG C), while for the direct effect, they were approximately unbiased when the outcome did not influence its own missingness (mDAGs B and E). These results extend previous simulation studies performed in the context of estimating the average causal effect, where in the absence of effect modification, multiple imputation was shown to give approximately unbiased estimates of the parameter of interest when outcome did not cause its own missingness and high bias otherwise.(14, 20, 22). More broadly, as illustrated with the VAHCS analysis, our study example highlights the importance of considering the missingness mechanism when planning the analysis, specifically in making decisions around analytic method and the possible need for sensitivity analysis, as well as when interpreting findings of mediation analysis in the presence of multivariable missingness. In this regard, m-DAGs provide a clear and transparent tool for depicting and assessing assumptions about the causes of missing data.

In our study, for the fully conditional specification multiple imputation approaches, whether and which interaction terms were incorporated in the imputation model did not have a material impact on the bias of the interventional [in]direct effect point estimates. This was in contrast with previous research, which has shown that appropriate inclusion of interaction terms to ensure the imputation model is approximately compatible with the analysis model can be important for preventing bias.(14, 35) Our observation might be explained by our simple data generation models that only included two-way interaction terms selected on the basis of content expertise, with their coefficient values set to values observed in the VAHCS dataset that were not very strong.

We found that for all multiple imputation approaches, BootMI returned model SEs with lower bias than MIBoot. The CIs had nominal coverage when using BootMI, but not MIBoot, under missingness mechanisms and multiple imputation methods where the effect estimates had little bias, and the imputation models included all analysis variables. This observation was in line with findings of Bartlett and Hughes (17), where in the presence of incompatibility or imputation model misspecification, MIBoot gave CIs that

under- or over-covered, while two versions of the BootMI approach returned CIs with nominal coverage. Bartlett and Hughes evaluated a “Boot MI percentile” approach as well as the modification developed by von Hippel and Bartlett (25) and referred to as BootMI in our paper. In the present paper, we did not investigate the performance of “Boot MI percentile”, which Bartlett and Hughes have found to be comparable to the von Hippel approach, because it is more computationally expensive.(17) In the context of the natural mediation effects, previously Wu and Jia had demonstrated that when using the so-called “product method” with normally distributed outcome and mediator variables, multivariate normal MIBoot performed well, but its performance deteriorated when the normality assumption was violated or when mediator and outcome influenced their own missingness.(13) The paper did not directly compare MIBoot with the Boot MI percentile approach but found that MIBoot had lower CI coverage when compared to a full information maximum likelihood nested within bootstrapping, an approach they speculated to have a similar performance to Boot MI percentile approach under the normality assumption. Despite this finding, the authors recommended using MIBoot when using multiple imputation to handle missing data, because the Boot MI percentile approach is computationally intensive. Fortunately, the von Hippel approach is a lot less computationally intensive and offers good statistical efficiency for a large number of bootstrap samples (e.g., 200) and imputations as few as two.(17) The R package bootImpute implements the von Hippel approach,(30) rendering it accessible to researchers.

Our simulation study was based on a real study and missing data were generated under seven missingness mechanisms that can be typically encountered in epidemiologic studies. These were depicted by m-DAGs, which provide a transparent framework for specifying assumptions about underlying missingness mechanisms in the context of mediation analysis. However, our m-DAGs were not exhaustive, and our findings may not be generalizable to all possible missingness mechanisms. We also only considered one complete data generation scenario with parameter values based on the VAHCS data, and no exposure-confounder interactions in mediator and outcome generation models. Future work assessing more complex settings, including in the presence of effect modification would be valuable. To reduce computational time in our simulation study, we implemented the simulation-based g-computation approach for estimating the interventional effects using 50 Monte Carlo draws, which is fewer than what would be used in practice to minimize Monte Carlo error. However, we do not expect this choice to have affected our simulation findings given estimates were obtained by averaging over 2,000 simulated dataset.

## Conclusion

In this paper we addressed three issues regarding the handling of multivariable missing data when estimating interventional [in]direct effects in the presence of intermediate confounding using g-computation. Our findings illustrate that a standard multiple imputation approach can be expected to perform well in terms of bias and variance reduction under missingness mechanisms in which the intermediate confounder, mediator and outcome variables do not directly influence their own missingness. This highlights the importance of substantive consideration of the missingness mechanism when planning and interpreting mediation analysis in the presence of multivariable missingness. Further, our results reinforce previous advice that to avoid bias in multiple imputation estimates the imputation model must be compatible with the substantive analysis. This can be achieved by using MI-SMCFCS or by incorporating in the imputation model all analysis variables used in the mediation analysis, which in this context include the intermediate confounder, mediator and outcome variables. Finally, our results suggest that BootMI (bootstrap followed by multiple imputation) is the preferred approach for obtaining standard errors and confidence intervals in the context of mediation analysis.

Figure 1. Directed acyclic graph (DAG) used in the data generation for the simulation study, based on the VAHCS case study
Note: Abbreviations: w, wave; CMD, common mental disorders

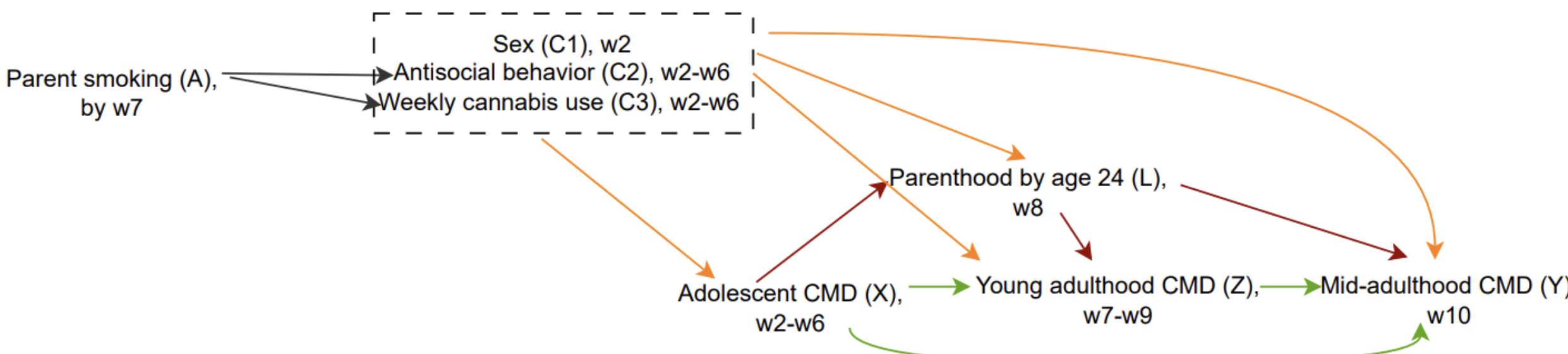

Figure 2. missingness-Directed acyclic graphs (m-DAGs) depicting the considered missingness mechanisms.
Note: To simplify the m-DAGs, the covariates with missing data ($C_2$,$C_3$) are presented as a single node with one missingness indicator ($\mathbf{M_C}$). When there is an arrow from $C_2$,$C_3$ to $\mathbf{M_C}$ (m-DAGs B to G) we assumed that each of the variables $C_2$ and $C_3$ cause their own missingness. The m-DAGs have been modified from Moreno-Betancur et al.[20]

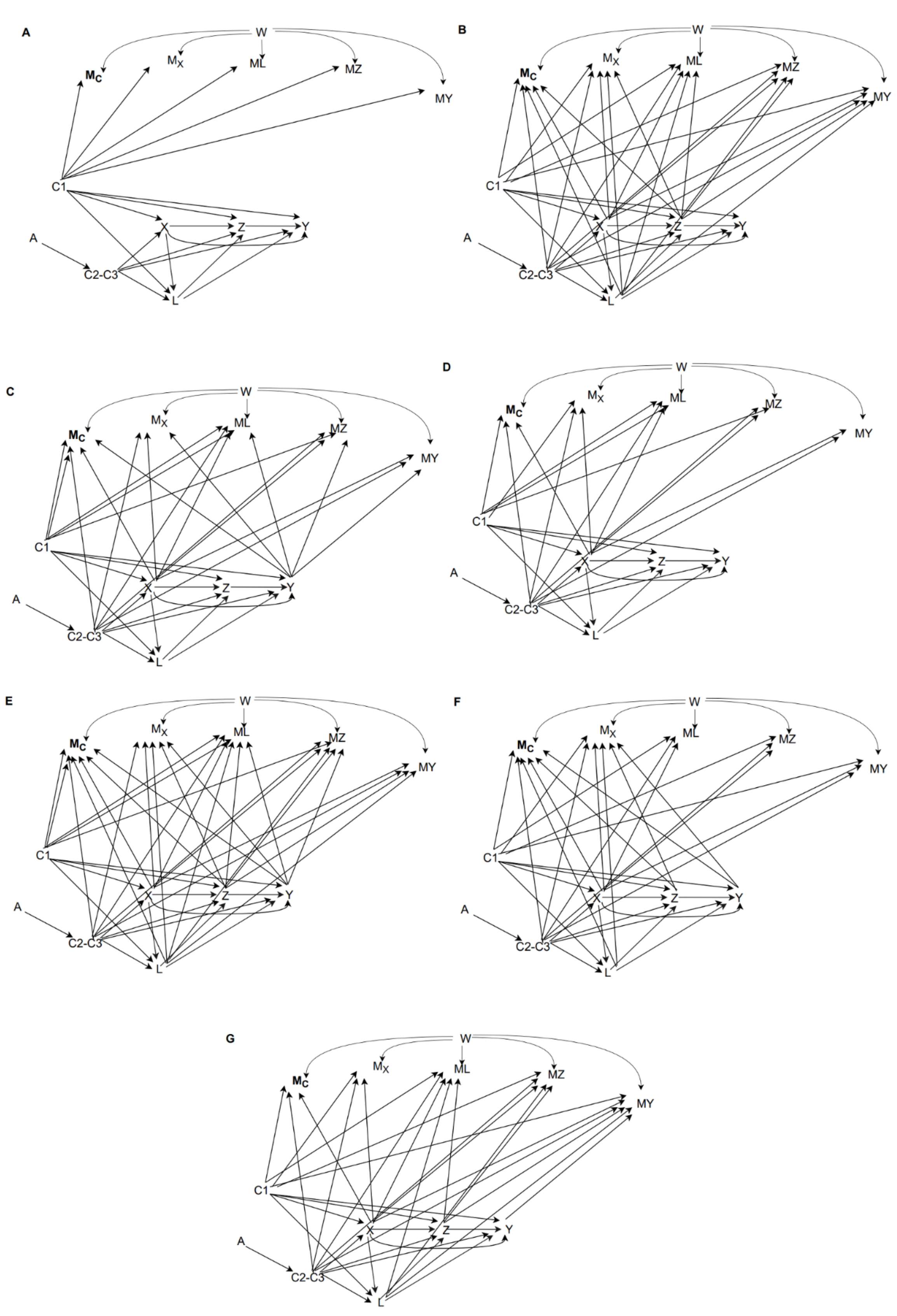

Figure 3. Relative bias (%) in estimates of the **interventional indirect effect** across the m-DAGs using Monte Carlo simulation-based g-computation as substantive mediation analysis method and different approaches to handle missing data, with MI methods implemented using the BootMI approach.
Note: See Table 2 for description of each of the approaches used for handling missing data. Due to perfect prediction, the MI-SMCFCS (BootMI) approach failed in 13 out of the 2,000 simulated datasets under m-DAG B and 7 out of 2,000 under E. The Monte Carlo standard errors ranged from 0.16% to 0.67% (eTable 4).

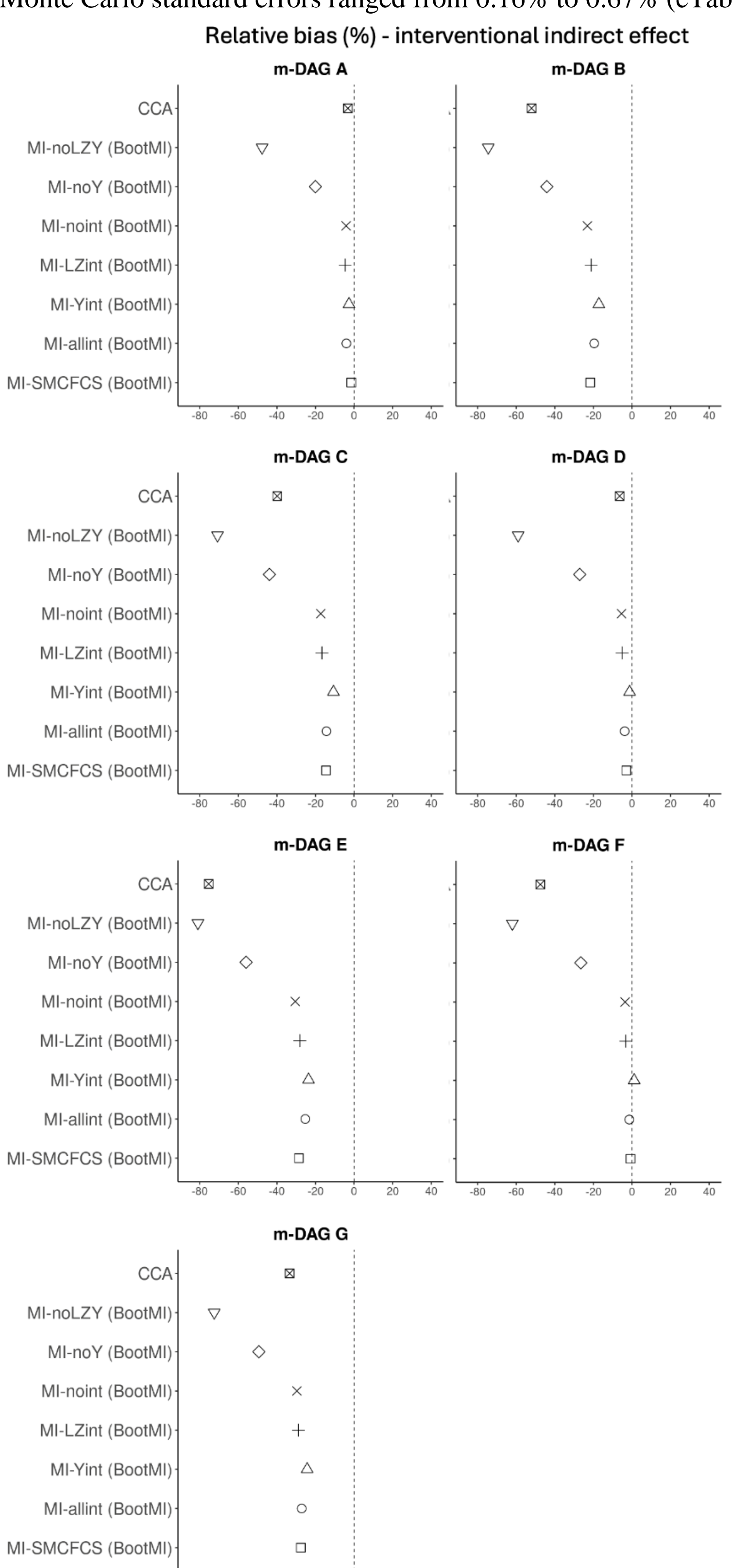

Figure 4. Percentage error in model-based standard error relative to empirical standard error in estimates of the **interventional indirect effect** across the m-DAGs using Monte Carlo simulation-based g-computation as substantive mediation analysis method and different approaches to handle missing data.

Note: See Table 2 for description of each of the approaches used for handling missing data. Due to perfect prediction, the MI-SMCFCS (BootMI) approach failed in 13 out of the 2,000 simulated datasets under m-DAG B and 7 out of 2,000 under E. The Monte Carlo standard errors ranged from 1.19% to 2.52% (eTable 4).

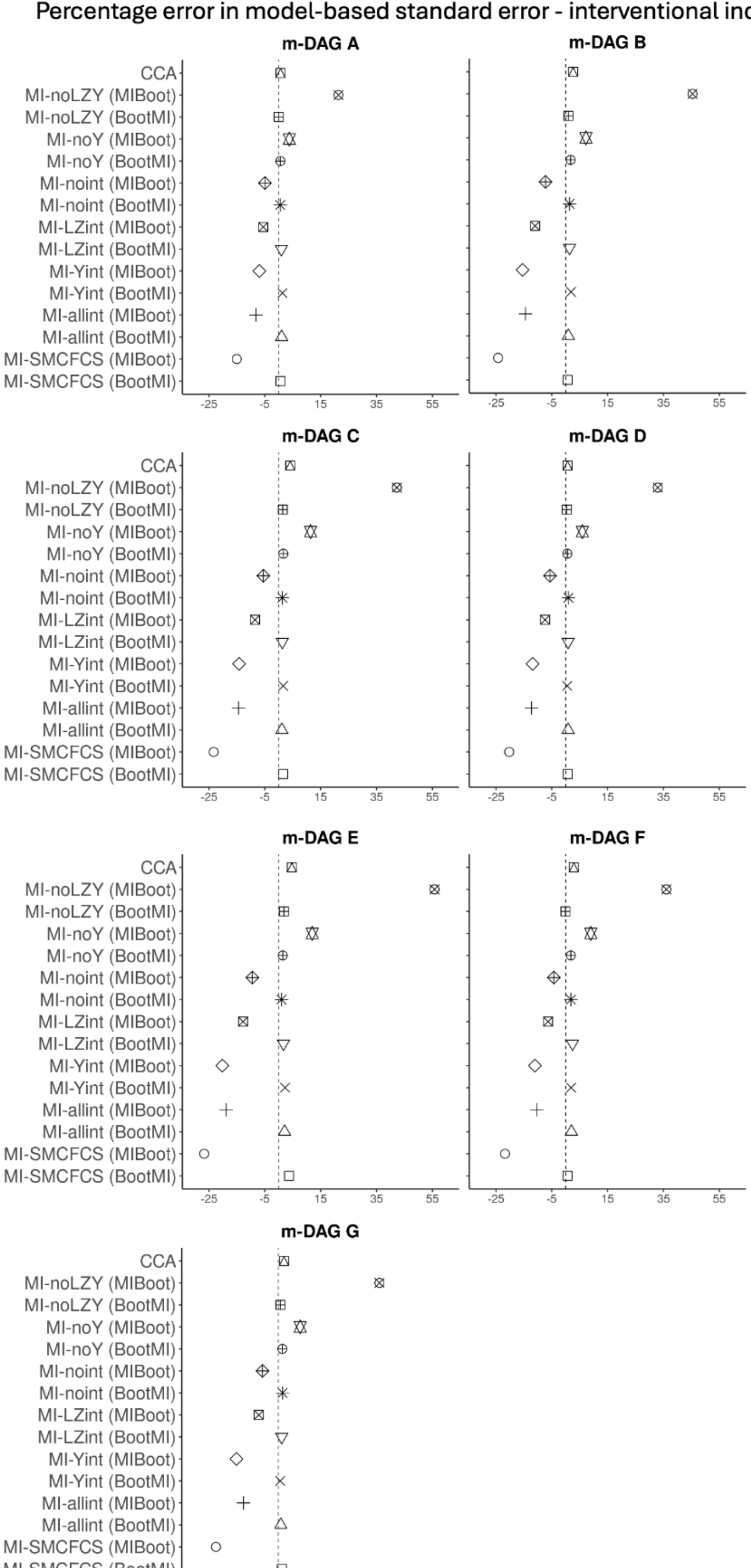

Table 1. Description of variables (all binary) considered for the case study, together with their distribution (number and percentage) and proportion with missing data in the VAHCS analytic sample, which excluded 20 participants who died during the follow-up (n= 1,923)

| | Variable | Type<br>Grouping/unit | Notation | N (% [a])<br>coded 1 | % with<br>missing data |
|---|---|---|---|---|---|
| Auxiliary variable | Parent smoking | 0= No by wave 7<br>1= Yes by wave 7 | $A$ | 712 (39) | 5 |
| Confounders | Sex assigned at birth | 0=Male<br>1= Female | $C1$ | 994 (52) | 0 |
| | Frequent cannabis use at baseline | 0=Less than weekly use across all waves 2-6<br>1=At least weekly use at any wave across waves 2-6 | $C2$ | 220 (17) [b] | 34 |
| | Antisocial behavior at baseline | 0=No across all waves 2-6<br>1=Yes at any wave across waves 2-6 | $C3$ | 341 (25) [b] | 28 |
| Exposure | CMD<br>(adolescence) | 0= total CIS score ≤11 or missing on more than 1 of waves 2-6<br>1= total CIS score >11 on at least 2 of waves 2-6 | $X$ | 438 (28) [b] | 20 |
| Intermediate confounder | Parenthood by age 24<br>(young adulthood) | 0= No by wave 8<br>1= Yes by wave 8 | $L$ | 131 (9) | 22 |
| Mediator | CMD<br>(young adulthood) | 0= total CIS score ≤11 or missing on more than 1 of waves 7-9<br>1= total CIS score >11 on at least 2 of waves 7-9 | $Z$ | 214 (14) [b] | 22 |
| Outcome | CMD<br>(mid-adulthood) | 0= total CIS score ≤11 at wave 10<br>1= total CIS score >11 at wave 10 | $Y$ | 261 (18.0) | 25 |
| | | | | With any missing data | 52 |

Abbreviations: N number, CIS Clinical Interview Schedule. [a] Proportions reported among those with observed data for the variable. [b] The numbers coded=0 are slightly different from the original investigation due to different decisions in coding (see eTable 2).

Table 2. Description of the approaches used for handling missing data in the simulation study

| Label | Description |
| --- | --- |
| CCA | Complete-case analysis |
| MI-noLZY [a] | The univariate imputation models were based on logistic regression models, incorporating only exposure ($X$) and pre-exposure confounders ($C_1, C_2, C_3$) and using main-effects terms (i.e., $L$, $Z$, $Y$, and interaction terms were not included as predictors in the univariate imputation models). Also, the auxiliary variable $A$ was included as a predictor in univariate models for imputing $C_2$ and $C_3$ (see Figure 2). |
| MI-noY [a] | The univariate imputation models were based on logistic regression models, incorporating only pre-exposure confounders ($C_1, C_2, C_3$), exposure ($X$), intermediate confounder ($L$), and mediator ($Z$) and using main-effects terms (i.e., $Y$ and interaction terms were not included as predictors in the univariate imputation models). Also, the auxiliary variable $A$ was included as a predictor in univariate models for imputing $C_2$ and $C_3$. |
| MI-noint | The univariate imputation models were based on logistic regression models incorporating all variables in the target analysis (i.e., $C_1, C_2, C_3, X, L, Z, Y$) but with main-effects terms only (i.e., no interaction terms were included as predictors in the univariate imputation models). Also, the auxiliary variable $A$ was included as a predictor in univariate models for imputing $C_2$ and $C_3$. |
| MI-LZint | The univariate imputation models were as for *MI-noint*, but additionally included, in the relevant univariate models, interaction terms so that the imputation model was approximately compatible with the models for the intermediate confounder and the mediator. |
| MI-Yint | The univariate imputation models were as for *MI-noint*, but additionally included, in the relevant univariate models, interaction terms so that the imputation model was approximately compatible with the outcome model. |
| MI-allint | The univariate imputation models were as for *MI-noint*, but additionally included, in the relevant univariate models, all the interaction terms included in *MI-LZint,* and *MI-Yint*. |
| MI-SMCFCS | Substantive-model compatibility was enforced for the outcome model, using the correctly specified outcome model, as used in the mediation analysis. Also, the auxiliary variable $A$ was included as a predictor in univariate models for imputing $C_2$ and $C_3$ |

[a] MI-noLZY and MI-noY clearly diverged from best-practice recommendation that the imputation models must incorporate all variables in the target analysis, in particular the outcome, and were only considered to demonstrate the extent of expected biases.

Box 1. General description of the steps involved in the Monte Carlo (MC) simulation-based g-computation approach as described in reference (3) used for estimating the interventional indirect and direct effects with binary exposure, mediator and outcome, in the presence of a binary intermediate confounder

1. Fit a regression model [a] for $L$ on $X$, and $\boldsymbol{C}$. Include exposure-confounder, and confounder-confounder interactions and non-linear terms as relevant in the regression model.

2. Fit a regression model [a] for $Z$ on $X, L,$ and $\boldsymbol{C}$. Include exposure-confounder, exposure-intermediate confounder, and confounder-confounder interactions and non-linear terms as relevant in the regression model.

3. Fit a regression model [a] for $Y$ on $X, L, Z$, and and $\boldsymbol{C}$. Include exposure-confounder, exposure-intermediate confounder, exposure-mediator, and confounder-confounder interactions and non-linear terms as relevant in the regression model.

4. Using Monte Carlo simulation, for each individual repeatedly perform random draws from the estimated distributions in Steps 1 and 2, setting $X = 1$ many times [b] and use the model in step 3 to predict the outcome when setting $X = 1$ and $L$ and $Z$ to the randomly drawn values. Obtain $\sum_{\boldsymbol{c}}\sum_{l}\sum_{z} \mathrm{P}(Y|X = 1, Z = z, L = l, \boldsymbol{C} = \boldsymbol{c})\mathrm{P}(Z = z, L = l|\boldsymbol{C} = \boldsymbol{c})\mathrm{P}(\boldsymbol{C} = \boldsymbol{c})$ as the average of the predicted outcomes across all individuals and Monte Carlo draws.

5. Repeat step 4 but perform random draws setting $X = 0$ and predict the outcome when setting $X = 0$ and $L$ and $Z$ to the randomly drawn values. Obtain $\sum_{\boldsymbol{c}}\sum_{l}\sum_{z} \mathrm{P}(Y|X = 0, Z = z, L = l, \boldsymbol{C} = \boldsymbol{c})\mathrm{P}(Z = z, L = l|\boldsymbol{C} = \boldsymbol{c})\mathrm{P}(\boldsymbol{C} = \boldsymbol{c})$ as the average of the predicted outcomes across all individuals and Monte Carlo draws.

6. Repeat step 4, but perform random draws from the estimated distribution in Step 1, setting $X = 1$, and from the estimated distribution in Step 2, setting $X = 0$, and predict the outcome when setting $X = 1$ and $L$ and $Z$ to the randomly drawn values. Obtain $\sum_{\boldsymbol{c}}\sum_{l}\sum_{z} P(Y|X = 1, Z = z, L = l, \boldsymbol{C} = \boldsymbol{c})\mathrm{P}(Z = z|X = 0, \boldsymbol{C} = \boldsymbol{c})\,\mathrm{P}(L = l|X = 1, \boldsymbol{C} = \boldsymbol{c})\mathrm{P}(\boldsymbol{C} = \boldsymbol{c})$ as the average of the predicted outcomes across all individuals and Monte Carlo draws.

7. Estimate the interventional indirect using formula (1), and the interventional direct effect using formula (2).

8. Obtain standard errors and confidence intervals using bootstrap.

[a] The approach relies on correctly specified models for the intermediate confounder(s), mediator, and outcome.

[b] To reduce computational time, for the simulation study, we used 50 Monte Carlo draws.

Box 2. Steps involved in the two approaches investigated for variance estimation in mediation analysis using bootstrap and multiple imputation

**Approach 1: MIBoot** (using Rubin's rules) (36)

1. Multiply impute missing data to generate M imputed datasets.

2. Analyze each imputed dataset to obtain a point estimate $\hat{\varphi}_{\mathrm{m}}$ for $\mathrm{m} = 1, \ldots, \mathrm{M}$ for target parameter $\varphi$ (either the indirect or direct interventional effect).

3. For each imputed dataset, draw B bootstrap samples with replacement, estimate the bootstrap variance ($\widehat{Var}_{\mathrm{bs}}(\hat{\varphi}_{\mathrm{m}})$, for $\mathrm{m} = 1, \ldots, \mathrm{M}$),

4. Apply Rubin's rules to pool the results across the imputed datasets to obtain the final MI point and variance estimates.

**Approach 2: BootMI** (von Hippel and Bartlett approach) (17, 25)

1. Draw B bootstrap samples with replacement from the data and multiply impute each bootstrap sample M times.

2. Analyze the imputed datasets, obtaining $\hat{\varphi}_{\mathrm{b,m}}$ for $\mathrm{m} = 1, \ldots, \mathrm{M}$ and $\mathrm{b} = 1, \ldots, \mathrm{B}$. The point estimate ($\bar{\varphi}_{\mathrm{BM}}$) is given by averaging first the estimates of the target parameter across the M imputed datasets for each bootstrap sample ($\bar{\varphi}_{\mathrm{b}} = \mathrm{M}^{-1} \sum_{\mathrm{m}=1}^{\mathrm{M}} \hat{\varphi}_{\mathrm{b,m}}$ for $\mathrm{b} = 1, \ldots, \mathrm{B}$), then averaging these across the B bootstrap samples ($\bar{\varphi}_{\mathrm{BM}} = \mathrm{B}^{-1} \sum_{\mathrm{b}=1}^{\mathrm{B}} \bar{\varphi}_{\mathrm{b}}$).

3. Estimate the variance using the method-of-moments estimator $\mathrm{Var}(\bar{\varphi}_{\mathrm{BM}}) = \left(1 + \frac{1}{\mathrm{B}}\right) \sigma_{\infty}^2 + \frac{1}{\mathrm{BM}} \sigma_{\mathrm{wb}}^2$, where $\hat{\sigma}_{\infty}^2$ and $\hat{\sigma}_{\mathrm{wb}}^2$ are estimated from the mean sum of squares within and between bootstraps (MSW and MSB), obtained from fitting a one-way analysis of variance (ANOVA) to the $\hat{\varphi}_{\mathrm{b,m}}$s: $\hat{\sigma}_{\infty}^2 = \frac{\mathrm{MSB}-\mathrm{MSW}}{\mathrm{M}}$ and $\hat{\sigma}_{\mathrm{wb}}^2 = \mathrm{MSW}$.

**eTable 1. Parameter values used to simulate the data and missingness under different missingness mechanisms considered**

| | | Variable | Intercept | A | C1 | C2 | C3 | X | L | Z | C1C3 | XL | XZ | Y | W |
|---|---|---|---|---|---|---|---|---|---|---|---|---|---|---|---|
| | | A | -0.45 | | | | | | | | | | | | |
| | | C1 | 0.10 | | | | | | | | | | | | |
| | | C2 | -1.92 | 0.63 | | | | | | | | | | | |
| | | C3 | -1.40 | 0.59 | | | | | | | | | | | |
| | | X | -2.23 | | 1.46 | 0.54 | 0.91 | | | | 0.28 | | | | |
| | | L | -3.12 | | 0.44 | -0.29 | 0.53 | 0.64 | | | 0.39 | | | | |
| | | Z | -2.75 | | 0.33 | -0.09 | -0.15 | 1.56 | -0.30 | | 0.93 | 0.23 | | | |
| Complete data | | Y | -2.05 | | 0.05 | 0.29 | 0.51 | 0.38 | 0.72 | 1.37 | -0.81 | 0.28 | 0.08 | | |
| | | W | -1.1 | | | | | | | | | | | | |
| | | MC2 | -2.9 | | 0.9 | | | | | | | | | | 7.5 |
| | | MC3 | -3.8 | | 0.9 | | | | | | | | | | 6 |
| | | MX | -4.6 | | 0.9 | | | | | | | | | | 4.5 |
| | | ML | -5.80 | | 0.9 | | | | | | | | | | 6.50 |
| | | MZ | -2.5 | | 0.9 | | | | | | | | | | 1.5 |
| | m-DAG A | MY | -2.7 | | 0.9 | | | | | | | | | | 2.7 |
| | | MC2 | -3.80 | | 0.9 | 0.9 | | 0.9 | 0.9 | 0.9 | | | | | 7.5 |
| | | MC3 | -4.80 | | 0.9 | | 0.9 | 0.9 | 0.9 | 0.9 | | | | | 6.5 |
| | | MX | -6.20 | | 0.9 | 0.9 | 0.9 | 0.9 | 0.9 | 0.9 | | | | | 5.5 |
| | | ML | -6.50 | | 0.9 | 0.9 | 0.9 | 0.9 | 0.9 | 0.9 | | | | | 6.50 |
| | | MZ | -3.50 | | 0.9 | 0.9 | 0.9 | 0.9 | 0.9 | 0.9 | | | | | 1.50 |
| | m-DAG B | MY | -3.40 | | 0.9 | 0.9 | 0.9 | 0.9 | 0.9 | 0.9 | | | | | 2.00 |
| | | MC2 | -3.80 | | 0.9 | 0.9 | | 0.9 | | | | | | 0.9 | 7.50 |
| | | MC3 | -4.70 | | 0.9 | | 0.9 | 0.9 | | | | | | 0.9 | 6.20 |
| | | MX | -6.20 | | 0.9 | 0.9 | 0.9 | 0.9 | | | | | | 0.9 | 5.50 |
| | | ML | -6.50 | | 0.9 | 0.9 | 0.9 | 0.9 | | | | | | 0.9 | 6.50 |
| | | MZ | -3.45 | | 0.9 | 0.9 | 0.9 | 0.9 | | | | | | 0.9 | 1.50 |
| | m-DAG C | MY | -3.45 | | 0.9 | 0.9 | 0.9 | 0.9 | | | | | | 0.9 | 2.35 |
| | | MC2 | -3.5 | | 0.9 | 0.9 | | 0.9 | | | | | | | 7.5 |
| | | MC3 | -4.5 | | 0.9 | | 0.9 | 0.9 | | | | | | | 6.5 |
| Imposing missing data | | MX | -6.00 | | 0.9 | 0.9 | 0.9 | 0.9 | | | | | | | 5.50 |
| | | ML | -6.30 | | 0.9 | 0.9 | 0.9 | 0.9 | | | | | | | 6.50 |
| | | MZ | -3.25 | | 0.9 | 0.9 | 0.9 | 0.9 | | | | | | | 1.5 |
| | m-DAG D | MY | -3.35 | | 0.9 | 0.9 | 0.9 | 0.9 | | | | | | | 2.60 |
| | | MC2 | -4.10 | | 0.9 | 0.9 | | 0.9 | 0.9 | 0.9 | | | | 0.9 | 7.50 |
| | | MC3 | -5.00 | | 0.9 | | 0.9 | 0.9 | 0.9 | 0.9 | | | | 0.9 | 6.50 |
| | | MX | -6.40 | | 0.9 | 0.9 | 0.9 | 0.9 | 0.9 | 0.9 | | | | 0.9 | 5.50 |
| | | ML | -6.70 | | 0.9 | 0.9 | 0.9 | 0.9 | 0.9 | 0.9 | | | | 0.9 | 6.50 |
| | | MZ | -3.70 | | 0.9 | 0.9 | 0.9 | 0.9 | 0.9 | 0.9 | | | | 0.9 | 1.40 |
| | m-DAG E | MY | -3.40 | | 0.9 | 0.9 | 0.9 | 0.9 | 0.9 | 0.9 | | | | | 2.00 |
| | | MC2 | -4.10 | | 0.9 | 0.9 | | 0.9 | 0.9 | 0.9 | | | | 0.9 | 7.50 |
| | | MC3 | -5.05 | | 0.9 | | 0.9 | 0.9 | 0.9 | 0.9 | | | | 0.9 | 6.50 |
| | | MX | -6.40 | | 0.9 | 0.9 | 0.9 | 0.9 | 0.9 | 0.9 | | | | 0.9 | 5.50 |
| | | ML | -6.30 | | 0.9 | 0.9 | 0.9 | 0.9 | | | | | | | 6.50 |
| | | MZ | -3.20 | | 0.9 | 0.9 | 0.9 | 0.9 | | | | | | | 1.40 |
| | m-DAG F | MY | -3.25 | | 0.9 | 0.9 | 0.9 | 0.9 | | | | | | | 2.30 |
| | | MC2 | -3.5 | | 0.9 | 0.9 | | 0.9 | | | | | | | 7.5 |
| | | MC3 | -4.5 | | 0.9 | | 0.9 | 0.9 | | | | | | | 6.5 |
| | | MX | -6.00 | | 0.9 | 0.9 | 0.9 | 0.9 | | | | | | | 5.50 |
| | | ML | -6.50 | | 0.9 | 0.9 | 0.9 | 0.9 | 0.9 | 0.9 | | | | | 6.50 |
| | | MZ | -3.50 | | 0.9 | 0.9 | 0.9 | 0.9 | 0.9 | 0.9 | | | | | 1.40 |
| | m-DAG G | MY | -3.50 | | 0.9 | 0.9 | 0.9 | 0.9 | 0.9 | 0.9 | | | | 0.9 | 2.40 |

**eTable 2. Description of variables in the simulated data (averaged over 2000 simulations)**

| | Variable | Notation | % coded 1 | % with missing data |
|---|---|---|---|---|
| Covariates | Sex assigned at birth | C1 | 53 | 0 |
| | Cannabis use [a] | C2 | 16 | 31 |
| | Antisocial behaviour [a] | C3 | 24 | 26 |
| Exposure | CMD in adolescence [b] | X | 28 | 16 |
| | Parenthood by age 24 | L | 8 | 19 |
| Mediator | CMD in young adulthood* | Z | 14 | 19 |
| Outcome | CMD in mid-adulthood | Y | 18 | 23 |
| Auxilliary variable | Parent smoking | A | | 0 |
| | | With any missing data | | 49 |

coded=0 for frequent cannabis use (n=1,052), antisocial behavior (n=1,035), and young adult CMD (n=1,277) were different from what was reported in the original investigation (n=1566, n=1675, and n=1,532 respectively) because of who was considered as having missing data. Here, for frequent cannabis use and antisocial behavior across waves 2-6, participants were recorded as missing if they had missing data for at least one wave and were coded=0 across other waves with available data. In the original investigation, participants were recorded as missing if they had missing data across all waves 2-6. For young adulthood CMD, here participants were coded as missing if they were not coded=1 for at least two waves and had missing data for two or more waves across waves 7-9. Again, in the original investigation, participants were recorded as missing if they had missing data across all waves 7-9. [b] The number coded=1 for adolescent CMD (n=438) was different than what was reported in the original investigation (n=316).(6) Here individuals were set to 1 if they had total CIS score >11 on 2 or more waves, regardless of missingness in other waves, and 0 if they had CIS score ≤11 across all waves 2 to 6 or had one wave of missing data for CIS score across all waves but had CIS score ≤11 across all other waves. In the original investigation all individuals with at least one wave of missing data across waves 2 to 6 for CIS

**eTable 3. The variables and interaction terms included in each imputation model for a multiple imputation approach that included all two-way interactions**

| | Variable imputed | Variables included in imputation model | | | | | | | | | | | | | | | | | |
|---|---|---|---|---|---|---|---|---|---|---|---|---|---|---|---|---|---|---|---|
| | | A | C1 | C2 | C3 | X | L | Z | Y | C1C3 | XL | XZ | XY | LC1 | LC3 | ZC1 | ZC3 | YC1 | YC3 |
| | C2 | 1 | 1 | | 1 | 1 | | | | | | | | | | | | | |
| | C3 | 1 | 1 | 1 | | 1 | | | | | | | | | | | | | |
| | X | | 1 | 1 | 1 | | | | | | | | | | | | | | |
| | L | | 1 | 1 | 1 | 1 | | | | | | | | | | | | | |
| | Z | | 1 | 1 | 1 | 1 | | | | | | | | | | | | | |
| noLZY | Y | | 1 | 1 | 1 | 1 | | | | | | | | | | | | | |
| | C2 | 1 | 1 | | 1 | 1 | 1 | 1 | | | | | | | | | | | |
| | C3 | 1 | 1 | 1 | | 1 | 1 | 1 | | | | | | | | | | | |
| | X | | 1 | 1 | 1 | | 1 | 1 | | | | | | | | | | | |
| | L | | 1 | 1 | 1 | 1 | | 1 | | | | | | | | | | | |
| | Z | | 1 | 1 | 1 | 1 | 1 | | | | | | | | | | | | |
| noY | Y | | 1 | 1 | 1 | 1 | 1 | 1 | | | | | | | | | | | |
| | C2 | 1 | 1 | | 1 | 1 | 1 | 1 | 1 | | | | | | | | | | |
| | C3 | 1 | 1 | 1 | | 1 | 1 | 1 | 1 | | | | | | | | | | |
| | X | | 1 | 1 | 1 | | 1 | 1 | 1 | | | | | | | | | | |
| | L | | 1 | 1 | 1 | 1 | | 1 | 1 | | | | | | | | | | |
| | Z | | 1 | 1 | 1 | 1 | 1 | | 1 | | | | | | | | | | |
| noint | Y | | 1 | 1 | 1 | 1 | 1 | 1 | | | | | | | | | | | |
| | C2 | 1 | 1 | | 1 | 1 | 1 | 1 | 1 | 1 | 1 | 1 | | 1 | 1 | 1 | 1 | | |
| | C3 | 1 | 1 | 1 | | 1 | 1 | 1 | 1 | | 1 | 1 | | 1 | | 1 | | | |
| | X | | 1 | 1 | 1 | | 1 | 1 | 1 | 1 | | | | 1 | 1 | 1 | 1 | | |
| | L | | 1 | 1 | 1 | | | 1 | 1 | 1 | | 1 | | | | 1 | 1 | | |
| | Z | | 1 | 1 | 1 | 1 | 1 | | 1 | 1 | 1 | | | 1 | 1 | | | | |
| LZint | Y | | 1 | 1 | 1 | 1 | 1 | 1 | | 1 | 1 | 1 | | 1 | 1 | 1 | 1 | | |
| | C2 | 1 | 1 | | 1 | 1 | 1 | 1 | 1 | 1 | 1 | 1 | 1 | | | | | 1 | 1 |
| | C3 | 1 | 1 | 1 | | 1 | 1 | 1 | 1 | | 1 | 1 | 1 | | | | | 1 | |
| | X | | 1 | 1 | 1 | | 1 | 1 | 1 | 1 | | | | | | | | 1 | 1 |
| | L | | 1 | 1 | 1 | 1 | | 1 | 1 | 1 | | 1 | 1 | | | | | 1 | 1 |
| | Z | | 1 | 1 | 1 | 1 | 1 | | 1 | 1 | 1 | | 1 | | | | | 1 | 1 |
| Yint | Y | | 1 | 1 | 1 | 1 | 1 | 1 | | 1 | 1 | 1 | | | | | | | 1 |
| | C2 | 1 | 1 | | 1 | 1 | 1 | 1 | 1 | 1 | 1 | 1 | 1 | 1 | 1 | 1 | 1 | 1 | 1 |
| | C3 | 1 | 1 | 1 | | 1 | 1 | 1 | 1 | | 1 | 1 | 1 | 1 | | 1 | | 1 | |
| | X | | 1 | 1 | 1 | | 1 | 1 | 1 | 1 | | | | 1 | 1 | 1 | 1 | 1 | 1 |
| | L | | 1 | 1 | 1 | 1 | | 1 | 1 | 1 | | 1 | 1 | | | 1 | 1 | 1 | 1 |
| | Z | | 1 | 1 | 1 | 1 | 1 | | 1 | 1 | 1 | | 1 | 1 | 1 | | | 1 | 1 |
| allint | Y | | 1 | 1 | 1 | 1 | 1 | 1 | | 1 | 1 | 1 | | 1 | 1 | 1 | 1 | | 1 |

**eTable 4A. Performance of the Monte Carlo simulation-based g-computation approach as substantive mediation analysis method and different approaches to handle missing data across the m-DAGs for estimates of the indirect effect**

| | | | INDIRECT EFFECT | | | | | | | | | | |
|---|---|---|---|---|---|---|---|---|---|---|---|---|---|
| | | Estimated risk | Abosulute bias | | Relative bias (%) | | Coverage (%) | | Empirical standard error | | % error in model standard error | | Number of |
| Causal diagram | Analysis | difference | Point estimate | MC SE | Point estimate | MC SE | Point estimat | MC SE | Point estimate | MC SE | Point estimate | MC SE | failed analysesd |
| | MI-SMCFCS (BootMI) | 0.0597 | -0.0009 | 0.0003 | -1.42 | 0.4977 | 94.40 | 0.5141 | 1.3481 | 0.0213 | 0.6190 | 1.6137 | 0 |
| | MI-SMCFCS (MIBoot) | 0.0600 | -0.0006 | 0.0003 | -0.98 | 0.5078 | 89.90 | 0.6738 | 1.3753 | 0.0218 | -15.0467 | 1.3591 | 0 |
| | MI-allint (BootMI) | 0.0581 | -0.0024 | 0.0003 | -4.00 | 0.4543 | 92.90 | 0.5743 | 1.2304 | 0.0195 | 1.0248 | 1.6208 | 0 |
| | MI-allint (MIBoot) | 0.0584 | -0.0022 | 0.0003 | -3.62 | 0.4596 | 90.95 | 0.6415 | 1.2448 | 0.0197 | -8.1358 | 1.4682 | 0 |
| | MI-Yint (BootMI) | 0.0590 | -0.0016 | 0.0003 | -2.62 | 0.4536 | 93.55 | 0.5493 | 1.2287 | 0.0194 | 1.3305 | 1.6252 | 0 |
| | MI-Yint (MIBoot) | 0.0592 | -0.0014 | 0.0003 | -2.33 | 0.4549 | 91.60 | 0.6203 | 1.2321 | 0.0195 | -6.9416 | 1.4871 | 0 |
| | MI-LZint (BootMI) | 0.0578 | -0.0028 | 0.0003 | -4.61 | 0.4406 | 92.85 | 0.5761 | 1.1934 | 0.0189 | 0.9389 | 1.6195 | 0 |
| A | MI-LZint (MIBoot) | 0.0579 | -0.0027 | 0.0003 | -4.40 | 0.4469 | 91.45 | 0.6253 | 1.2105 | 0.0191 | -5.5736 | 1.5088 | 0 |
| | MI-noint (BootMI) | 0.0581 | -0.0025 | 0.0003 | -4.12 | 0.4407 | 93.20 | 0.5629 | 1.1936 | 0.0189 | 0.5272 | 1.6121 | 0 |
| | MI-noint (MIBoot) | 0.0580 | -0.0025 | 0.0003 | -4.18 | 0.4445 | 91.65 | 0.6186 | 1.2040 | 0.0190 | -4.9614 | 1.5189 | 0 |
| | MI-noY (BootMI) | 0.0484 | -0.0121 | 0.0002 | -20.06 | 0.3819 | 73.40 | 0.9880 | 1.0344 | 0.0164 | 0.5858 | 1.6153 | 0 |
| | MI-noY (MIBoot) | 0.0483 | -0.0123 | 0.0002 | -20.25 | 0.3851 | 75.35 | 0.9637 | 1.0431 | 0.0165 | 3.7970 | 1.6599 | 0 |
| | MI-noLZY (BootMI) | 0.0317 | -0.0289 | 0.0002 | -47.70 | 0.2729 | 6.50 | 0.5512 | 0.7391 | 0.0117 | -0.0087 | 1.6082 | 0 |
| | MI-noLZY (MIBoot) | 0.0316 | -0.0289 | 0.0002 | -47.76 | 0.2767 | 11.80 | 0.7214 | 0.7494 | 0.0119 | 21.3716 | 1.9418 | 0 |
| | CCA | 0.0586 | -0.0019 | 0.0004 | -3.19 | 0.5830 | 93.90 | 0.5352 | 1.5791 | 0.0250 | 0.6499 | 1.6232 | 0 |
| | MI-SMCFCS (BootMI) | 0.0475 | -0.0131 | 0.0003 | -21.65 | 0.5567 | 81.38 | 0.8733 | 1.5029 | 0.0238 | 0.6969 | 1.6283 | 13 |
| | MI-SMCFCS (MIBoot) | 0.0479 | -0.0127 | 0.0003 | -20.94 | 0.5716 | 68.90 | 1.0351 | 1.5481 | 0.0245 | -24.2158 | 1.2207 | 0 |
| | MI-allint (BootMI) | 0.0487 | -0.0119 | 0.0003 | -19.64 | 0.4827 | 80.50 | 0.8859 | 1.3075 | 0.0207 | 0.9397 | 1.6264 | 0 |
| | MI-allint (MIBoot) | 0.0493 | -0.0113 | 0.0003 | -18.65 | 0.4882 | 74.55 | 0.9740 | 1.3222 | 0.0209 | -14.4103 | 1.3745 | 0 |
| | MI-Yint (BootMI) | 0.0502 | -0.0104 | 0.0003 | -17.19 | 0.4875 | 83.35 | 0.8330 | 1.3205 | 0.0209 | 1.8973 | 1.6425 | 0 |
| | MI-Yint (MIBoot) | 0.0510 | -0.0095 | 0.0003 | -15.72 | 0.5036 | 77.40 | 0.9352 | 1.3639 | 0.0216 | -15.4503 | 1.3577 | 0 |
| | MI-LZint (BootMI) | 0.0477 | -0.0128 | 0.0003 | -21.18 | 0.4516 | 76.60 | 0.9467 | 1.2233 | 0.0193 | 1.3493 | 1.6342 | 0 |
| B | MI-LZint (MIBoot) | 0.0482 | -0.0124 | 0.0003 | -20.49 | 0.4645 | 72.00 | 1.0040 | 1.2582 | 0.0199 | -10.9769 | 1.4290 | 0 |
| | MI-noint (BootMI) | 0.0465 | -0.0140 | 0.0003 | -23.14 | 0.4340 | 72.45 | 0.9990 | 1.1756 | 0.0186 | 1.3606 | 1.6363 | 0 |
| | MI-noint (MIBoot) | 0.0467 | -0.0138 | 0.0003 | -22.82 | 0.4425 | 68.60 | 1.0378 | 1.1986 | 0.0190 | -7.2291 | 1.4898 | 0 |
| | MI-noY (BootMI) | 0.0338 | -0.0268 | 0.0002 | -44.20 | 0.3343 | 21.90 | 0.9248 | 0.9055 | 0.0143 | 1.7600 | 1.6494 | 0 |
| | MI-noY (MIBoot) | 0.0338 | -0.0267 | 0.0002 | -44.14 | 0.3459 | 26.35 | 0.9851 | 0.9368 | 0.0148 | 7.2347 | 1.7237 | 0 |
| | MI-noLZY (BootMI) | 0.0154 | -0.0451 | 0.0001 | -74.53 | 0.1924 | 0.00 | 0.0000 | 0.5211 | 0.0082 | 0.9721 | 1.6501 | 0 |
| | MI-noLZY (MIBoot) | 0.0154 | -0.0452 | 0.0001 | -74.63 | 0.1959 | 0.00 | 0.0000 | 0.5306 | 0.0084 | 45.3920 | 2.3391 | 0 |
| | CCA | 0.0290 | -0.0315 | 0.0003 | -52.08 | 0.5208 | 47.20 | 1.1163 | 1.4105 | 0.0223 | 2.6483 | 1.7067 | 0 |
| | MI-SMCFCS (BootMI) | 0.0518 | -0.0088 | 0.0003 | -14.52 | 0.5645 | 87.75 | 0.7331 | 1.5291 | 0.0242 | 1.5272 | 1.6352 | 0 |
| | MI-SMCFCS (MIBoot) | 0.0524 | -0.0081 | 0.0004 | -13.42 | 0.5834 | 77.75 | 0.9300 | 1.5802 | 0.0250 | -23.3091 | 1.2337 | 0 |
| | MI-allint (BootMI) | 0.0519 | -0.0087 | 0.0003 | -14.33 | 0.4910 | 86.90 | 0.7545 | 1.3300 | 0.0210 | 1.1208 | 1.6283 | 0 |
| | MI-allint (MIBoot) | 0.0522 | -0.0083 | 0.0003 | -13.75 | 0.5015 | 80.85 | 0.8799 | 1.3583 | 0.0215 | -14.4209 | 1.3727 | 0 |
| | MI-Yint (BootMI) | 0.0541 | -0.0065 | 0.0003 | -10.69 | 0.4969 | 90.10 | 0.6678 | 1.3459 | 0.0213 | 1.5796 | 1.6342 | 0 |
| | MI-Yint (MIBoot) | 0.0550 | -0.0056 | 0.0003 | -9.25 | 0.5067 | 85.30 | 0.7918 | 1.3725 | 0.0217 | -14.1987 | 1.3752 | 0 |
| | MI-LZint (BootMI) | 0.0505 | -0.0101 | 0.0003 | -16.66 | 0.4499 | 83.40 | 0.8320 | 1.2186 | 0.0193 | 1.2839 | 1.6293 | 0 |
| C | MI-LZint (MIBoot) | 0.0508 | -0.0097 | 0.0003 | -16.09 | 0.4637 | 80.25 | 0.8902 | 1.2560 | 0.0199 | -8.4914 | 1.4664 | 0 |
| | MI-noint (BootMI) | 0.0501 | -0.0105 | 0.0003 | -17.26 | 0.4413 | 82.35 | 0.8525 | 1.1954 | 0.0189 | 1.2927 | 1.6312 | 0 |
| | MI-noint (MIBoot) | 0.0502 | -0.0103 | 0.0003 | -17.08 | 0.4487 | 79.40 | 0.9043 | 1.2155 | 0.0192 | -5.5169 | 1.5149 | 0 |
| | MI-noY (BootMI) | 0.0340 | -0.0266 | 0.0002 | -43.89 | 0.3267 | 21.35 | 0.9163 | 0.8850 | 0.0140 | 1.6224 | 1.6460 | 0 |
| | MI-noY (MIBoot) | 0.0340 | -0.0266 | 0.0002 | -43.92 | 0.3386 | 26.20 | 0.9832 | 0.9172 | 0.0145 | 11.3881 | 1.7879 | 0 |
| | MI-noLZY (BootMI) | 0.0177 | -0.0429 | 0.0001 | -70.80 | 0.2055 | 0.00 | 0.0000 | 0.5566 | 0.0088 | 1.4262 | 1.6488 | 0 |
| | MI-noLZY (MIBoot) | 0.0177 | -0.0428 | 0.0001 | -70.74 | 0.2137 | 0.10 | 0.0707 | 0.5788 | 0.0092 | 42.2740 | 2.2858 | 0 |
| | CCA | 0.0364 | -0.0242 | 0.0003 | -39.88 | 0.5432 | 66.10 | 1.0585 | 1.4713 | 0.0233 | 4.1008 | 1.7025 | 0 |
| | MI-SMCFCS (BootMI) | 0.0588 | -0.0018 | 0.0003 | -2.94 | 0.5557 | 94.55 | 0.5076 | 1.5051 | 0.0238 | 0.7044 | 1.6172 | 0 |
| | MI-SMCFCS (MIBoot) | 0.0592 | -0.0014 | 0.0003 | -2.31 | 0.5717 | 87.10 | 0.7495 | 1.5486 | 0.0245 | -20.2453 | 1.2781 | 0 |
| | MI-allint (BootMI) | 0.0583 | -0.0023 | 0.0003 | -3.80 | 0.4905 | 93.65 | 0.5453 | 1.3285 | 0.0210 | 0.8694 | 1.6182 | 0 |
| | MI-allint (MIBoot) | 0.0585 | -0.0021 | 0.0003 | -3.40 | 0.5033 | 89.85 | 0.6753 | 1.3632 | 0.0216 | -12.2540 | 1.4035 | 0 |
| | MI-Yint (BootMI) | 0.0598 | -0.0008 | 0.0003 | -1.25 | 0.4972 | 94.15 | 0.5248 | 1.3468 | 0.0213 | 0.5137 | 1.6137 | 0 |
| | MI-Yint (MIBoot) | 0.0603 | -0.0003 | 0.0003 | -0.44 | 0.5059 | 90.60 | 0.6525 | 1.3703 | 0.0217 | -11.8703 | 1.4097 | 0 |
| | MI-LZint (BootMI) | 0.0575 | -0.0031 | 0.0003 | -5.12 | 0.4658 | 93.35 | 0.5571 | 1.2615 | 0.0200 | 0.8549 | 1.6198 | 0 |
| D | MI-LZint (MIBoot) | 0.0577 | -0.0028 | 0.0003 | -4.68 | 0.4752 | 90.60 | 0.6525 | 1.2872 | 0.0204 | -7.3861 | 1.4812 | 0 |
| | MI-noint (BootMI) | 0.0573 | -0.0033 | 0.0003 | -5.45 | 0.4574 | 92.60 | 0.5853 | 1.2390 | 0.0196 | 0.9415 | 1.6211 | 0 |
| | MI-noint (MIBoot) | 0.0574 | -0.0032 | 0.0003 | -5.31 | 0.4662 | 91.00 | 0.6399 | 1.2628 | 0.0200 | -5.6464 | 1.5087 | 0 |
| | MI-noY (BootMI) | 0.0441 | -0.0164 | 0.0002 | -27.14 | 0.3754 | 58.95 | 1.1000 | 1.0168 | 0.0161 | 0.5829 | 1.6202 | 0 |
| | MI-noY (MIBoot) | 0.0441 | -0.0165 | 0.0002 | -27.23 | 0.3869 | 63.45 | 1.0768 | 1.0479 | 0.0166 | 5.9481 | 1.6965 | 0 |
| | MI-noLZY (BootMI) | 0.0248 | -0.0358 | 0.0001 | -59.05 | 0.2449 | 0.55 | 0.1654 | 0.6632 | 0.0105 | 0.3810 | 1.6206 | 0 |
| | MI-noLZY (MIBoot) | 0.0248 | -0.0358 | 0.0002 | -59.07 | 0.2520 | 2.70 | 0.3624 | 0.6826 | 0.0108 | 32.9450 | 2.1301 | 0 |
| | CCA | 0.0567 | -0.0039 | 0.0004 | -6.42 | 0.6755 | 93.30 | 0.5591 | 1.8297 | 0.0289 | 0.6670 | 1.6303 | 0 |
| | MI-SMCFCS (BootMI) | 0.0433 | -0.0173 | 0.0003 | -28.57 | 0.5606 | 74.76 | 0.9730 | 1.5158 | 0.0240 | 3.6684 | 1.6803 | 7 |
| | MI-SMCFCS (MIBoot) | 0.0438 | -0.0167 | 0.0004 | -27.62 | 0.5905 | 59.75 | 1.0966 | 1.5993 | 0.0253 | -26.6952 | 1.1873 | 0 |
| | MI-allint (BootMI) | 0.0453 | -0.0153 | 0.0003 | -25.28 | 0.4915 | 73.65 | 0.9851 | 1.3313 | 0.0211 | 2.0802 | 1.6512 | 0 |
| | MI-allint (MIBoot) | 0.0458 | -0.0148 | 0.0003 | -24.37 | 0.5094 | 63.85 | 1.0743 | 1.3799 | 0.0218 | -18.8040 | 1.3086 | 0 |
| | MI-Yint (BootMI) | 0.0463 | -0.0143 | 0.0003 | -23.60 | 0.5011 | 75.25 | 0.9650 | 1.3572 | 0.0215 | 2.2541 | 1.6565 | 0 |
| | MI-Yint (MIBoot) | 0.0470 | -0.0136 | 0.0003 | -22.39 | 0.5280 | 66.35 | 1.0566 | 1.4302 | 0.0226 | -20.2030 | 1.2869 | 0 |
| | MI-LZint (BootMI) | 0.0435 | -0.0170 | 0.0003 | -28.11 | 0.4484 | 66.10 | 1.0585 | 1.2145 | 0.0192 | 1.6782 | 1.6460 | 0 |
| E | MI-LZint (MIBoot) | 0.0440 | -0.0166 | 0.0003 | -27.38 | 0.4651 | 59.80 | 1.0963 | 1.2598 | 0.0199 | -12.7799 | 1.4054 | 0 |
| | MI-noint (BootMI) | 0.0421 | -0.0185 | 0.0003 | -30.49 | 0.4328 | 60.30 | 1.0941 | 1.1723 | 0.0185 | 0.9761 | 1.6356 | 0 |
| | MI-noint (MIBoot) | 0.0423 | -0.0182 | 0.0003 | -30.13 | 0.4445 | 55.30 | 1.1117 | 1.2040 | 0.0190 | -9.4460 | 1.4595 | 0 |
| | MI-noY (BootMI) | 0.0267 | -0.0339 | 0.0002 | -56.00 | 0.2963 | 5.35 | 0.5032 | 0.8027 | 0.0127 | 1.4384 | 1.6538 | 0 |
| | MI-noY (MIBoot) | 0.0266 | -0.0340 | 0.0002 | -56.09 | 0.3091 | 7.60 | 0.5926 | 0.8373 | 0.0132 | 11.9828 | 1.8083 | 0 |
| | MI-noLZY (BootMI) | 0.0115 | -0.0490 | 0.0001 | -80.96 | 0.1637 | 0.00 | 0.0000 | 0.4433 | 0.0070 | 1.8603 | 1.6811 | 0 |
| | MI-noLZY (MIBoot) | 0.0115 | -0.0491 | 0.0001 | -81.00 | 0.1704 | 0.00 | 0.0000 | 0.4617 | 0.0073 | 55.7667 | 2.5193 | 0 |
| | CCA | 0.0149 | -0.0457 | 0.0002 | -75.38 | 0.3928 | 11.45 | 0.7120 | 1.0638 | 0.0168 | 4.6762 | 1.7947 | 0 |
| | MI-SMCFCS (BootMI) | 0.0601 | -0.0005 | 0.0004 | -0.80 | 0.5789 | 94.10 | 0.5269 | 1.5681 | 0.0248 | 0.6527 | 1.6164 | 0 |
| | MI-SMCFCS (MIBoot) | 0.0608 | 0.0002 | 0.0004 | 0.33 | 0.5985 | 86.60 | 0.7617 | 1.6210 | 0.0256 | -21.7371 | 1.2549 | 0 |
| | MI-allint (BootMI) | 0.0597 | -0.0009 | 0.0003 | -1.42 | 0.5004 | 93.95 | 0.5331 | 1.3554 | 0.0214 | 2.0444 | 1.6386 | 0 |
| | MI-allint (MIBoot) | 0.0602 | -0.0003 | 0.0003 | -0.58 | 0.5046 | 90.95 | 0.6415 | 1.3667 | 0.0216 | -10.3325 | 1.4342 | 0 |
| | MI-Yint (BootMI) | 0.0612 | 0.0007 | 0.0003 | 1.09 | 0.5046 | 94.60 | 0.5054 | 1.3668 | 0.0216 | 1.9433 | 1.6358 | 0 |
| | MI-Yint (MIBoot) | 0.0619 | 0.0014 | 0.0003 | 2.25 | 0.5124 | 91.35 | 0.6286 | 1.3879 | 0.0219 | -11.0203 | 1.4233 | 0 |
| | MI-LZint (BootMI) | 0.0586 | -0.0019 | 0.0003 | -3.20 | 0.4677 | 93.50 | 0.5512 | 1.2669 | 0.0200 | 2.4306 | 1.6445 | 0 |
| F | MI-LZint (MIBoot) | 0.0590 | -0.0016 | 0.0003 | -2.60 | 0.4796 | 91.10 | 0.6367 | 1.2991 | 0.0205 | -6.3127 | 1.4982 | 0 |
| | MI-noint (BootMI) | 0.0584 | -0.0022 | 0.0003 | -3.59 | 0.4619 | 93.35 | 0.5571 | 1.2510 | 0.0198 | 1.8431 | 1.6365 | 0 |
| | MI-noint (MIBoot) | 0.0585 | -0.0020 | 0.0003 | -3.37 | 0.4690 | 92.30 | 0.5961 | 1.2703 | 0.0201 | -4.2544 | 1.5315 | 0 |
| | MI-noY (BootMI) | 0.0445 | -0.0161 | 0.0002 | -26.54 | 0.3749 | 62.60 | 1.0820 | 1.0156 | 0.0161 | 1.8400 | 1.6409 | 0 |
| | MI-noY (MIBoot) | 0.0445 | -0.0161 | 0.0002 | -26.55 | 0.3829 | 66.60 | 1.0546 | 1.0371 | 0.0164 | 8.9729 | 1.7443 | 0 |
| | MI-noLZY (BootMI) | 0.0230 | -0.0376 | 0.0001 | -62.05 | 0.2351 | 0.20 | 0.0999 | 0.6369 | 0.0101 | -0.1019 | 1.6159 | 0 |
| | MI-noLZY (MIBoot) | 0.0230 | -0.0376 | 0.0001 | -62.00 | 0.2414 | 1.15 | 0.2384 | 0.6538 | 0.0103 | 36.0254 | 2.1806 | 0 |
| | CCA | 0.0318 | -0.0288 | 0.0003 | -47.53 | 0.5424 | 55.55 | 1.1111 | 1.4692 | 0.0232 | 2.9139 | 1.6915 | 0 |
| | MI-SMCFCS (BootMI) | 0.0439 | -0.0167 | 0.0003 | -27.60 | 0.5005 | 72.15 | 1.0023 | 1.3556 | 0.0214 | 1.3146 | 1.6355 | 0 |
| | MI-SMCFCS (MIBoot) | 0.0443 | -0.0163 | 0.0003 | -26.85 | 0.5193 | 59.90 | 1.0959 | 1.4066 | 0.0222 | -22.2789 | 1.2531 | 0 |
| | MI-allint (BootMI) | 0.0441 | -0.0164 | 0.0003 | -27.15 | 0.4441 | 68.15 | 1.0418 | 1.2029 | 0.0190 | 0.9409 | 1.6278 | 0 |
| | MI-allint (MIBoot) | 0.0446 | -0.0160 | 0.0003 | -26.34 | 0.4487 | 61.25 | 1.0894 | 1.2155 | 0.0192 | -12.4350 | 1.4073 | 0 |
| | MI-Yint (BootMI) | 0.0458 | -0.0147 | 0.0003 | -24.32 | 0.4559 | 72.85 | 0.9945 | 1.2348 | 0.0195 | 0.7337 | 1.6250 | 0 |
| | MI-Yint (MIBoot) | 0.0466 | -0.0140 | 0.0003 | -23.04 | 0.4695 | 67.00 | 1.0514 | 1.2717 | 0.0201 | -14.9462 | 1.3671 | 0 |
| | MI-LZint (BootMI) | 0.0431 | -0.0174 | 0.0002 | -28.81 | 0.4105 | 61.65 | 1.0873 | 1.1119 | 0.0176 | 1.2516 | 1.6349 | 0 |
| G | MI-LZint (MIBoot) | 0.0435 | -0.0171 | 0.0003 | -28.20 | 0.4172 | 56.80 | 1.1076 | 1.1301 | 0.0179 | -6.9460 | 1.4951 | 0 |
| | MI-noint (BootMI) | 0.0426 | -0.0180 | 0.0002 | -29.64 | 0.4004 | 59.20 | 1.0989 | 1.0846 | 0.0172 | 1.5823 | 1.6409 | 0 |
| | MI-noint (MIBoot) | 0.0428 | -0.0178 | 0.0002 | -29.34 | 0.4097 | 55.75 | 1.1106 | 1.1098 | 0.0176 | -5.6113 | 1.5165 | 0 |
| | MI-noY (BootMI) | 0.0307 | -0.0299 | 0.0002 | -49.34 | 0.3114 | 11.15 | 0.7038 | 0.8434 | 0.0133 | 1.5147 | 1.6472 | 0 |
| | MI-noY (MIBoot) | 0.0308 | -0.0298 | 0.0002 | -49.18 | 0.3226 | 13.85 | 0.7724 | 0.8737 | 0.0138 | 7.8797 | 1.7377 | 0 |
| | MI-noLZY (BootMI) | 0.0167 | -0.0439 | 0.0001 | -72.50 | 0.2018 | 0.00 | 0.0000 | 0.5465 | 0.0086 | 0.8182 | 1.6419 | 0 |
| | MI-noLZY (MIBoot) | 0.0165 | -0.0440 | 0.0001 | -72.70 | 0.2056 | 0.00 | 0.0000 | 0.5568 | 0.0088 | 36.2619 | 2.1954 | 0 |
| | CCA | 0.0403 | -0.0202 | 0.0004 | -33.42 | 0.5895 | 74.70 | 0.9721 | 1.5967 | 0.0253 | 2.0942 | 1.6717 | 0 |

**eTable 4B. Performance of the Monte Carlo simulation-based g-computation approach as substantive mediation analysis method and different approaches to handle missing data across the m-DAGs for estimates of the direct effect**

| | | DIRECT EFFECT | | | | | | | | | | | |
|---|---|---|---|---|---|---|---|---|---|---|---|---|---|
| | | Estimated risk | Abosulute bias | | Relative bias (%) | | Coverage (%) | | Empirical standard error | | % error in model standard error | | Number of |
| Causal diagram | Analysis | difference | Point estimate | MC SE | Point estimate | MC SE | Point estimate | MC SE | Point estimate | MC SE | Point estimate | MC SE | failed analysesd |
| | MI-SMCFCS (Bc | 0.0659 | -0.0001 | 0.0006 | -0.16 | 0.9272 | 94.20 | 0.5227 | 2.7368 | 0.0433 | -1.6836 | 1.5646 | 0 |
| | MI-SMCFCS (MI | 0.0652 | -0.0008 | 0.0006 | -1.20 | 0.9352 | 89.30 | 0.6912 | 2.7602 | 0.0437 | -15.9969 | 1.3329 | 0 |
| | MI-allint (Boot | 0.0646 | -0.0014 | 0.0006 | -2.18 | 0.9127 | 94.35 | 0.5163 | 2.6938 | 0.0426 | -1.5009 | 1.5675 | 0 |
| | MI-allint (MIBc | 0.0643 | -0.0017 | 0.0006 | -2.57 | 0.9265 | 90.00 | 0.6708 | 2.7346 | 0.0432 | -15.0867 | 1.3470 | 0 |
| | MI-Yint (BootM | 0.0630 | -0.0030 | 0.0006 | -4.48 | 0.9190 | 94.20 | 0.5227 | 2.7124 | 0.0429 | -1.9037 | 1.5615 | 0 |
| | MI-Yint (MIBoc | 0.0628 | -0.0032 | 0.0006 | -4.92 | 0.9244 | 89.55 | 0.6840 | 2.7282 | 0.0431 | -15.0768 | 1.3473 | 0 |
| | MI-LZint (Bootl | 0.0658 | -0.0002 | 0.0006 | -0.25 | 0.9062 | 94.55 | 0.5076 | 2.6746 | 0.0423 | -1.6749 | 1.5647 | 0 |
| A | MI-LZint (MIBo | 0.0656 | -0.0004 | 0.0006 | -0.54 | 0.9205 | 90.05 | 0.6693 | 2.7169 | 0.0430 | -14.1646 | 1.3617 | 0 |
| | MI-noint (Boot | 0.0663 | 0.0003 | 0.0006 | 0.53 | 0.9094 | 94.15 | 0.5248 | 2.6841 | 0.0425 | -2.0891 | 1.5580 | 0 |
| | MI-noint (MIBc | 0.0660 | 0.0000 | 0.0006 | -0.04 | 0.9141 | 90.60 | 0.6525 | 2.6981 | 0.0427 | -13.6193 | 1.3701 | 0 |
| | MI-noY (BootM | 0.0697 | 0.0037 | 0.0005 | 5.68 | 0.8300 | 94.15 | 0.5248 | 2.4497 | 0.0387 | -2.2871 | 1.5550 | 0 |
| | MI-noY (MIBoo | 0.0697 | 0.0037 | 0.0006 | 5.56 | 0.8410 | 93.90 | 0.5352 | 2.4822 | 0.0393 | -4.9073 | 1.5078 | 0 |
| | MI-noLZY (Boo | 0.0833 | 0.0173 | 0.0005 | 26.25 | 0.8292 | 89.95 | 0.6723 | 2.4473 | 0.0387 | -2.1255 | 1.5570 | 0 |
| | MI-noLZY (MIB | 0.0831 | 0.0171 | 0.0006 | 25.97 | 0.8429 | 89.20 | 0.6940 | 2.4877 | 0.0393 | -3.3265 | 1.5324 | 0 |
| | CCA | 0.0648 | -0.0012 | 0.0007 | -1.87 | 1.0745 | 95.45 | 0.4660 | 3.1714 | 0.0502 | -0.2838 | 1.5881 | 0 |
| | MI-SMCFCS (Bc | 0.0714 | 0.0054 | 0.0006 | 8.21 | 0.9759 | 95.42 | 0.4690 | 2.8711 | 0.0456 | 0.6157 | 1.6072 | 13 |
| | MI-SMCFCS (MI | 0.0705 | 0.0045 | 0.0007 | 6.79 | 0.9946 | 88.85 | 0.7038 | 2.9355 | 0.0464 | -18.6210 | 1.2920 | 0 |
| | MI-allint (Boot | 0.0700 | 0.0040 | 0.0006 | 6.14 | 0.9524 | 95.10 | 0.4827 | 2.8110 | 0.0445 | 0.5065 | 1.6006 | 0 |
| | MI-allint (MIBc | 0.0698 | 0.0038 | 0.0006 | 5.80 | 0.9659 | 89.90 | 0.6738 | 2.8510 | 0.0451 | -16.4663 | 1.3259 | 0 |
| | MI-Yint (BootM | 0.0678 | 0.0018 | 0.0006 | 2.80 | 0.9596 | 95.60 | 0.4586 | 2.8324 | 0.0448 | 0.8028 | 1.6050 | 0 |
| | MI-Yint (MIBoc | 0.0671 | 0.0011 | 0.0006 | 1.69 | 0.9726 | 89.85 | 0.6753 | 2.8706 | 0.0454 | -17.3060 | 1.3129 | 0 |
| | MI-LZint (Bootl | 0.0720 | 0.0060 | 0.0006 | 9.17 | 0.9416 | 95.10 | 0.4827 | 2.7793 | 0.0440 | 0.8583 | 1.6055 | 0 |
| B | MI-LZint (MIBo | 0.0718 | 0.0058 | 0.0006 | 8.76 | 0.9619 | 90.05 | 0.6693 | 2.8390 | 0.0449 | -15.8082 | 1.3363 | 0 |
| | MI-noint (Boot | 0.0708 | 0.0048 | 0.0006 | 7.33 | 0.9456 | 94.60 | 0.5054 | 2.7910 | 0.0441 | 0.6143 | 1.6019 | 0 |
| | MI-noint (MIBc | 0.0706 | 0.0047 | 0.0006 | 7.05 | 0.9589 | 89.90 | 0.6738 | 2.8301 | 0.0448 | -15.6279 | 1.3390 | 0 |
| | MI-noY (BootM | 0.0731 | 0.0071 | 0.0006 | 10.79 | 0.8419 | 94.75 | 0.4987 | 2.4849 | 0.0393 | 0.4708 | 1.5992 | 0 |
| | MI-noY (MIBoo | 0.0731 | 0.0071 | 0.0006 | 10.79 | 0.8543 | 93.75 | 0.5413 | 2.5216 | 0.0399 | -4.5102 | 1.5145 | 0 |
| | MI-noLZY (Boo | 0.0817 | 0.0157 | 0.0006 | 23.77 | 0.8369 | 91.70 | 0.6169 | 2.4701 | 0.0391 | 0.6171 | 1.6011 | 0 |
| | MI-noLZY (MIB | 0.0817 | 0.0157 | 0.0006 | 23.82 | 0.8490 | 90.70 | 0.6494 | 2.5058 | 0.0396 | -3.1500 | 1.5358 | 0 |
| | CCA | 0.0573 | -0.0087 | 0.0008 | -13.16 | 1.1537 | 93.60 | 0.5473 | 3.4051 | 0.0539 | 2.0017 | 1.6275 | 0 |
| | MI-SMCFCS (Bc | 0.0481 | -0.0179 | 0.0006 | -27.12 | 0.9310 | 87.95 | 0.7279 | 2.7479 | 0.0435 | 0.5178 | 1.6036 | 0 |
| | MI-SMCFCS (MI | 0.0466 | -0.0194 | 0.0006 | -29.34 | 0.9446 | 78.60 | 0.9171 | 2.7879 | 0.0441 | -19.0847 | 1.2870 | 0 |
| | MI-allint (Boot | 0.0454 | -0.0206 | 0.0006 | -31.23 | 0.9016 | 85.80 | 0.7805 | 2.6610 | 0.0421 | 0.1081 | 1.5967 | 0 |
| | MI-allint (MIBc | 0.0449 | -0.0211 | 0.0006 | -31.93 | 0.9002 | 78.85 | 0.9131 | 2.6569 | 0.0420 | -15.8358 | 1.3376 | 0 |
| | MI-Yint (BootM | 0.0422 | -0.0238 | 0.0006 | -36.13 | 0.9018 | 83.70 | 0.8259 | 2.6616 | 0.0421 | 0.7960 | 1.6080 | 0 |
| | MI-Yint (MIBoc | 0.0409 | -0.0250 | 0.0006 | -37.96 | 0.9132 | 72.25 | 1.0012 | 2.6953 | 0.0426 | -17.4359 | 1.3128 | 0 |
| | MI-LZint (Bootl | 0.0479 | -0.0181 | 0.0006 | -27.48 | 0.8764 | 87.15 | 0.7483 | 2.5867 | 0.0409 | 0.0223 | 1.5953 | 0 |
| C | MI-LZint (MIBo | 0.0474 | -0.0186 | 0.0006 | -28.15 | 0.8873 | 80.90 | 0.8790 | 2.6188 | 0.0414 | -14.0871 | 1.3649 | 0 |
| | MI-noint (Boot | 0.0481 | -0.0179 | 0.0006 | -27.13 | 0.8760 | 86.95 | 0.7532 | 2.5856 | 0.0409 | 0.0190 | 1.5952 | 0 |
| | MI-noint (MIBc | 0.0478 | -0.0182 | 0.0006 | -27.57 | 0.8975 | 80.95 | 0.8781 | 2.6490 | 0.0419 | -14.9666 | 1.3509 | 0 |
| | MI-noY (BootM | 0.0582 | -0.0078 | 0.0005 | -11.75 | 0.7973 | 92.70 | 0.5817 | 2.3533 | 0.0372 | -0.8224 | 1.5803 | 0 |
| | MI-noY (MIBoo | 0.0581 | -0.0079 | 0.0005 | -11.95 | 0.8106 | 91.95 | 0.6084 | 2.3925 | 0.0378 | -3.6202 | 1.5297 | 0 |
| | MI-noLZY (Boo | 0.0705 | 0.0045 | 0.0005 | 6.81 | 0.8066 | 94.50 | 0.5098 | 2.3807 | 0.0377 | -0.9553 | 1.5775 | 0 |
| | MI-noLZY (MIB | 0.0703 | 0.0043 | 0.0005 | 6.51 | 0.8209 | 93.85 | 0.5372 | 2.4228 | 0.0383 | -2.9709 | 1.5394 | 0 |
| | CCA | 0.0125 | -0.0535 | 0.0007 | -81.01 | 0.9914 | 57.40 | 1.1057 | 2.9260 | 0.0463 | 0.5172 | 1.6118 | 0 |
| | MI-SMCFCS (Bc | 0.0673 | 0.0014 | 0.0006 | 2.05 | 0.9648 | 94.95 | 0.4896 | 2.8476 | 0.0450 | 0.4185 | 1.5995 | 0 |
| | MI-SMCFCS (MI | 0.0663 | 0.0003 | 0.0006 | 0.42 | 0.9754 | 90.40 | 0.6587 | 2.8790 | 0.0455 | -16.9355 | 1.3189 | 0 |
| | MI-allint (Boot | 0.0648 | -0.0012 | 0.0006 | -1.82 | 0.9371 | 94.95 | 0.4896 | 2.7659 | 0.0437 | 0.9913 | 1.6082 | 0 |
| | MI-allint (MIBc | 0.0647 | -0.0013 | 0.0006 | -1.97 | 0.9505 | 90.60 | 0.6525 | 2.8054 | 0.0444 | -14.8861 | 1.3510 | 0 |
| | MI-Yint (BootM | 0.0623 | -0.0037 | 0.0006 | -5.62 | 0.9445 | 94.85 | 0.4942 | 2.7877 | 0.0441 | 0.8497 | 1.6059 | 0 |
| | MI-Yint (MIBoc | 0.0617 | -0.0043 | 0.0006 | -6.46 | 0.9576 | 89.40 | 0.6883 | 2.8264 | 0.0447 | -15.9921 | 1.3333 | 0 |
| | MI-LZint (Bootl | 0.0668 | 0.0008 | 0.0006 | 1.26 | 0.9241 | 95.20 | 0.4780 | 2.7275 | 0.0431 | 0.7458 | 1.6049 | 0 |
| D | MI-LZint (MIBo | 0.0663 | 0.0003 | 0.0006 | 0.45 | 0.9340 | 91.30 | 0.6302 | 2.7566 | 0.0436 | -13.1647 | 1.3780 | 0 |
| | MI-noint (Boot | 0.0680 | 0.0020 | 0.0006 | 3.06 | 0.9208 | 95.35 | 0.4708 | 2.7177 | 0.0430 | 0.9093 | 1.6065 | 0 |
| | MI-noint (MIBc | 0.0676 | 0.0016 | 0.0006 | 2.40 | 0.9375 | 91.10 | 0.6367 | 2.7671 | 0.0438 | -13.4604 | 1.3733 | 0 |
| | MI-noY (BootM | 0.0744 | 0.0084 | 0.0006 | 12.69 | 0.8370 | 94.80 | 0.4965 | 2.4705 | 0.0391 | 0.8124 | 1.6046 | 0 |
| | MI-noY (MIBoo | 0.0742 | 0.0082 | 0.0006 | 12.36 | 0.8483 | 93.70 | 0.5433 | 2.5038 | 0.0396 | -2.8075 | 1.5415 | 0 |
| | MI-noLZY (Boo | 0.0907 | 0.0247 | 0.0006 | 37.45 | 0.8436 | 84.85 | 0.8017 | 2.4898 | 0.0394 | 0.3451 | 1.5964 | 0 |
| | MI-noLZY (MIB | 0.0907 | 0.0247 | 0.0006 | 37.42 | 0.8583 | 83.95 | 0.8208 | 2.5331 | 0.0401 | -1.8347 | 1.5565 | 0 |
| | CCA | 0.0644 | -0.0016 | 0.0008 | -2.45 | 1.1743 | 95.00 | 0.4873 | 3.4659 | 0.0548 | 1.5132 | 1.6192 | 0 |
| | MI-SMCFCS (Bc | 0.0764 | 0.0104 | 0.0007 | 15.69 | 0.9921 | 95.03 | 0.4867 | 2.9230 | 0.0463 | 2.7180 | 1.6389 | 7 |
| | MI-SMCFCS (MI | 0.0753 | 0.0093 | 0.0007 | 14.03 | 1.0127 | 86.85 | 0.7557 | 2.9890 | 0.0473 | -18.6856 | 1.2912 | 0 |
| | MI-allint (Boot | 0.0741 | 0.0081 | 0.0006 | 12.34 | 0.9731 | 94.95 | 0.4896 | 2.8721 | 0.0454 | 1.8050 | 1.6208 | 0 |
| | MI-allint (MIBc | 0.0735 | 0.0075 | 0.0007 | 11.41 | 0.9888 | 88.70 | 0.7079 | 2.9183 | 0.0462 | -17.2569 | 1.3136 | 0 |
| | MI-Yint (BootM | 0.0723 | 0.0063 | 0.0006 | 9.58 | 0.9785 | 94.95 | 0.4896 | 2.8880 | 0.0457 | 1.9456 | 1.6231 | 0 |
| | MI-Yint (MIBoc | 0.0716 | 0.0056 | 0.0007 | 8.45 | 0.9986 | 87.95 | 0.7279 | 2.9475 | 0.0466 | -18.0918 | 1.3003 | 0 |
| | MI-LZint (Bootl | 0.0769 | 0.0109 | 0.0006 | 16.55 | 0.9618 | 94.00 | 0.5310 | 2.8389 | 0.0449 | 1.0258 | 1.6085 | 0 |
| E | MI-LZint (MIBo | 0.0764 | 0.0104 | 0.0006 | 15.70 | 0.9718 | 88.70 | 0.7079 | 2.8684 | 0.0454 | -15.4403 | 1.3419 | 0 |
| | MI-noint (Boot | 0.0760 | 0.0100 | 0.0006 | 15.18 | 0.9713 | 94.25 | 0.5205 | 2.8669 | 0.0453 | 0.4261 | 1.5987 | 0 |
| | MI-noint (MIBc | 0.0759 | 0.0099 | 0.0006 | 14.96 | 0.9837 | 88.15 | 0.7227 | 2.9034 | 0.0459 | -16.2745 | 1.3290 | 0 |
| | MI-noY (BootM | 0.0786 | 0.0126 | 0.0006 | 19.14 | 0.8388 | 93.60 | 0.5473 | 2.4757 | 0.0392 | 0.3030 | 1.5972 | 0 |
| | MI-noY (MIBoo | 0.0784 | 0.0124 | 0.0006 | 18.80 | 0.8589 | 91.95 | 0.6084 | 2.5350 | 0.0401 | -3.7716 | 1.5260 | 0 |
| | MI-noLZY (Boo | 0.0850 | 0.0190 | 0.0006 | 28.73 | 0.8437 | 89.80 | 0.6767 | 2.4903 | 0.0394 | -0.4095 | 1.5849 | 0 |
| | MI-noLZY (MIB | 0.0847 | 0.0187 | 0.0006 | 28.33 | 0.8521 | 88.90 | 0.7024 | 2.5150 | 0.0398 | -2.5918 | 1.5446 | 0 |
| | CCA | 0.0234 | -0.0426 | 0.0007 | -64.47 | 1.0614 | 72.05 | 1.0034 | 3.1329 | 0.0495 | 0.0477 | 1.6010 | 0 |
| | MI-SMCFCS (Bc | 0.0684 | 0.0024 | 0.0007 | 3.60 | 0.9988 | 94.90 | 0.4919 | 2.9480 | 0.0466 | -0.5607 | 1.5846 | 0 |
| | MI-SMCFCS (MI | 0.0672 | 0.0012 | 0.0007 | 1.79 | 1.0142 | 89.10 | 0.6968 | 2.9936 | 0.0473 | -19.1305 | 1.2845 | 0 |
| | MI-allint (Boot | 0.0661 | 0.0001 | 0.0006 | 0.20 | 0.9661 | 95.05 | 0.4850 | 2.8516 | 0.0451 | -0.3637 | 1.5869 | 0 |
| | MI-allint (MIBc | 0.0657 | -0.0003 | 0.0006 | -0.49 | 0.9836 | 90.20 | 0.6648 | 2.9032 | 0.0459 | -17.0573 | 1.3168 | 0 |
| | MI-Yint (BootM | 0.0634 | -0.0026 | 0.0006 | -3.88 | 0.9748 | 94.35 | 0.5163 | 2.8772 | 0.0455 | -0.4814 | 1.5847 | 0 |
| | MI-Yint (MIBoc | 0.0625 | -0.0034 | 0.0007 | -5.23 | 0.9887 | 88.65 | 0.7093 | 2.9180 | 0.0461 | -17.6656 | 1.3077 | 0 |
| | MI-LZint (Bootl | 0.0677 | 0.0017 | 0.0006 | 2.60 | 0.9577 | 95.00 | 0.4873 | 2.8266 | 0.0447 | -0.5749 | 1.5840 | 0 |
| F | MI-LZint (MIBo | 0.0673 | 0.0013 | 0.0006 | 1.95 | 0.9751 | 90.50 | 0.6556 | 2.8779 | 0.0455 | -15.9074 | 1.3348 | 0 |
| | MI-noint (Boot | 0.0686 | 0.0026 | 0.0006 | 3.99 | 0.9555 | 95.10 | 0.4827 | 2.8201 | 0.0446 | -0.4931 | 1.5847 | 0 |
| | MI-noint (MIBc | 0.0685 | 0.0025 | 0.0006 | 3.73 | 0.9679 | 90.10 | 0.6678 | 2.8567 | 0.0452 | -15.1441 | 1.3468 | 0 |
| | MI-noY (BootM | 0.0724 | 0.0064 | 0.0006 | 9.72 | 0.8366 | 94.65 | 0.5032 | 2.4692 | 0.0391 | -0.4103 | 1.5862 | 0 |
| | MI-noY (MIBoo | 0.0722 | 0.0062 | 0.0006 | 9.40 | 0.8536 | 94.15 | 0.5248 | 2.5193 | 0.0398 | -2.5334 | 1.5459 | 0 |
| | MI-noLZY (Boo | 0.0876 | 0.0216 | 0.0006 | 32.66 | 0.8411 | 86.90 | 0.7545 | 2.4825 | 0.0393 | -0.6554 | 1.5815 | 0 |
| | MI-noLZY (MIB | 0.0874 | 0.0214 | 0.0006 | 32.40 | 0.8613 | 86.90 | 0.7545 | 2.5422 | 0.0402 | -1.5800 | 1.5608 | 0 |
| | CCA | 0.0394 | -0.0266 | 0.0007 | -40.25 | 1.1234 | 86.50 | 0.7641 | 3.3156 | 0.0524 | 0.9235 | 1.6135 | 0 |
| | MI-SMCFCS (Bc | 0.0538 | -0.0122 | 0.0006 | -18.54 | 0.9002 | 91.05 | 0.6383 | 2.6570 | 0.0420 | 0.7059 | 1.6050 | 0 |
| | MI-SMCFCS (MI | 0.0527 | -0.0133 | 0.0006 | -20.18 | 0.9140 | 84.15 | 0.8166 | 2.6978 | 0.0427 | -17.3664 | 1.3130 | 0 |
| | MI-allint (Boot | 0.0512 | -0.0148 | 0.0006 | -22.42 | 0.8786 | 90.90 | 0.6431 | 2.5933 | 0.0410 | 0.6956 | 1.6040 | 0 |
| | MI-allint (MIBc | 0.0508 | -0.0152 | 0.0006 | -23.07 | 0.8933 | 82.80 | 0.8438 | 2.6366 | 0.0417 | -15.8204 | 1.3371 | 0 |
| | MI-Yint (BootM | 0.0488 | -0.0172 | 0.0006 | -26.13 | 0.8842 | 89.15 | 0.6954 | 2.6097 | 0.0413 | 0.8283 | 1.6071 | 0 |
| | MI-Yint (MIBoc | 0.0475 | -0.0185 | 0.0006 | -28.00 | 0.8964 | 79.60 | 0.9011 | 2.6459 | 0.0418 | -16.3984 | 1.3280 | 0 |
| | MI-LZint (Bootl | 0.0534 | -0.0126 | 0.0006 | -19.13 | 0.8616 | 91.45 | 0.6253 | 2.5429 | 0.0402 | 0.7957 | 1.6058 | 0 |
| G | MI-LZint (MIBo | 0.0526 | -0.0133 | 0.0006 | -20.23 | 0.8763 | 85.45 | 0.7884 | 2.5863 | 0.0409 | -13.8332 | 1.3683 | 0 |
| | MI-noint (Boot | 0.0537 | -0.0123 | 0.0006 | -18.62 | 0.8635 | 91.20 | 0.6335 | 2.5486 | 0.0403 | 0.1502 | 1.5963 | 0 |
| | MI-noint (MIBc | 0.0534 | -0.0126 | 0.0006 | -19.09 | 0.8723 | 86.65 | 0.7605 | 2.5746 | 0.0407 | -13.3618 | 1.3756 | 0 |
| | MI-noY (BootM | 0.0607 | -0.0053 | 0.0005 | -7.96 | 0.7905 | 94.80 | 0.4965 | 2.3331 | 0.0369 | 0.8855 | 1.6064 | 0 |
| | MI-noY (MIBoo | 0.0606 | -0.0054 | 0.0005 | -8.14 | 0.7963 | 93.85 | 0.5372 | 2.3504 | 0.0372 | -3.4393 | 1.5319 | 0 |
| | MI-noLZY (Boo | 0.0714 | 0.0054 | 0.0005 | 8.21 | 0.7942 | 95.30 | 0.4732 | 2.3442 | 0.0371 | 0.9542 | 1.6070 | 0 |
| | MI-noLZY (MIB | 0.0714 | 0.0054 | 0.0005 | 8.12 | 0.8082 | 93.75 | 0.5413 | 2.3854 | 0.0377 | -2.9200 | 1.5401 | 0 |
| | CCA | 0.0414 | -0.0246 | 0.0007 | -37.21 | 1.1195 | 87.85 | 0.7305 | 3.3042 | 0.0523 | 0.6614 | 1.6088 | 0 |

eFigure 1. Relative bias (%) in estimates of the interventional direct effect across the m-DAGs using Monte Carlo simulation-based g-computation as substantive mediation analysis method and different approaches to handle missing data, with MI methods implemented using the BootMI approach.

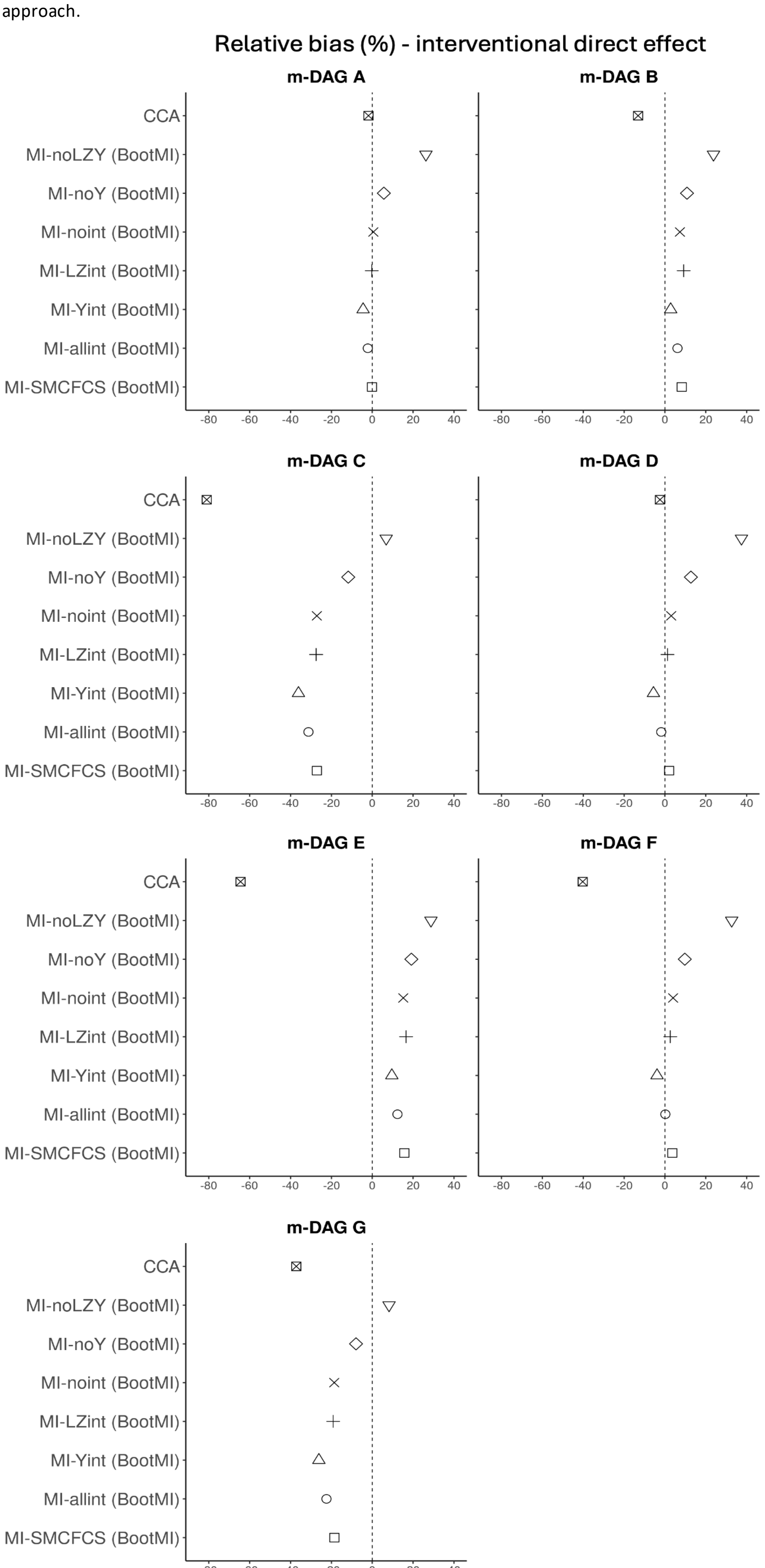

Note: See Table 2 for description of each of the approaches used for handling missing data. Due to perfect prediction, the MI-SMCFCS (BootMI) approach failed in 13 out of the 2,000 simulated datasets under m-DAG B and 7 out of 2,000 under E. The Monte Carlo standard errors ranged from 0.75% to 1.17% (eTable 4).

eFigure 2. Percentage error in model-based standard error relative to empirical standard error in estimates of the interventional direct effect across the m-DAGs using Monte Carlo simulation-based g-computation as substantive mediation analysis method and different approaches to handle missing data.

Percentage error in model-based standard error - interventional direct effect

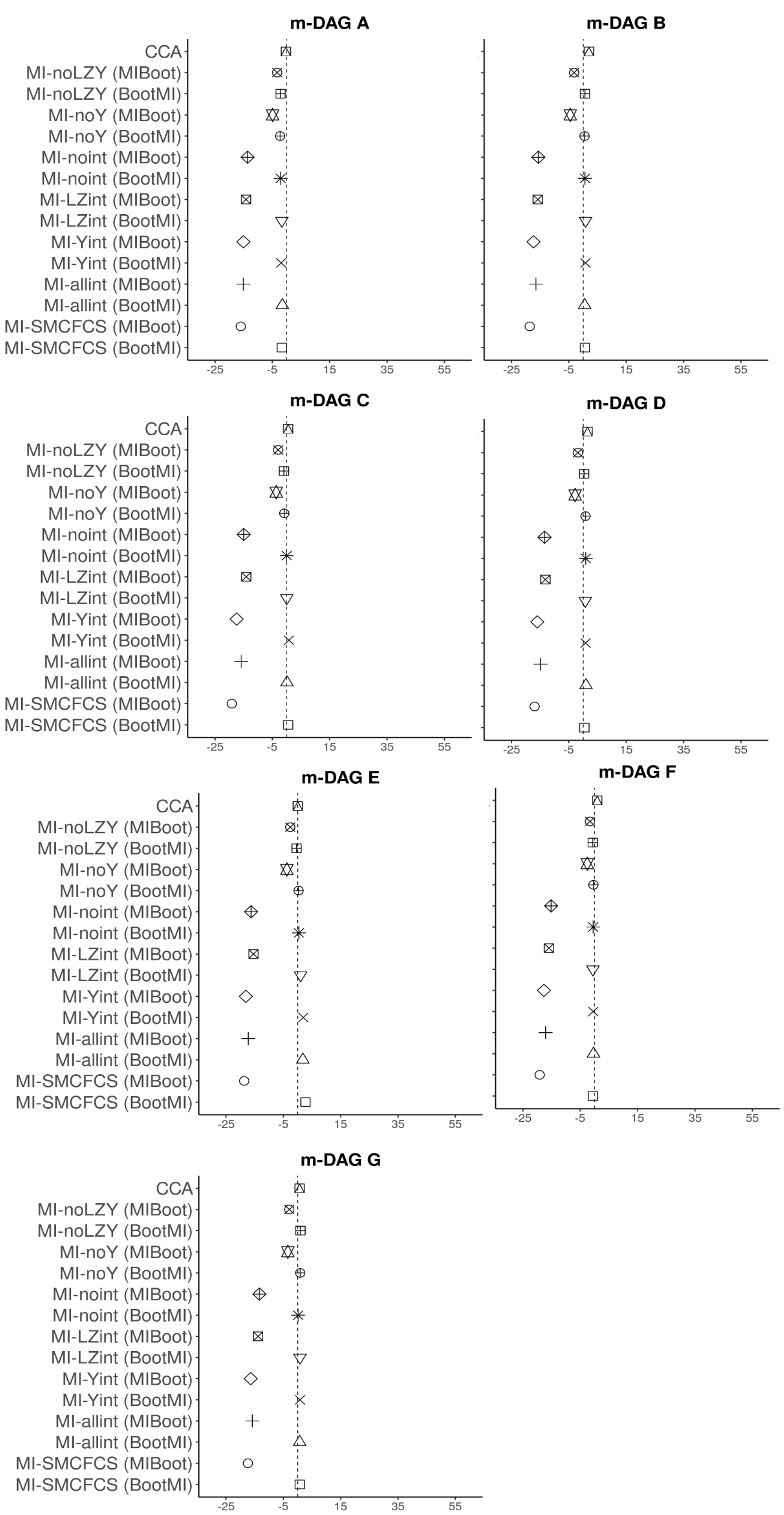

Note: See Table 2 for description of each of the approaches used for handling missing data. Due to perfect prediction, the MI-SMCFCS (BootMI) approach failed in 13 out of the 2,000 simulated datasets under m-DAG B and 7 out of 2,000 under E. The Monte Carlo standard errors ranged from 1.29% to 1.64% (eTable 4).

eFigure 3. Coverage probability (%) in estimates of the interventional indirect effect across the m-DAGs using Monte Carlo simulation-based g-computation as substantive mediation analysis method and different approaches to handle missing data.

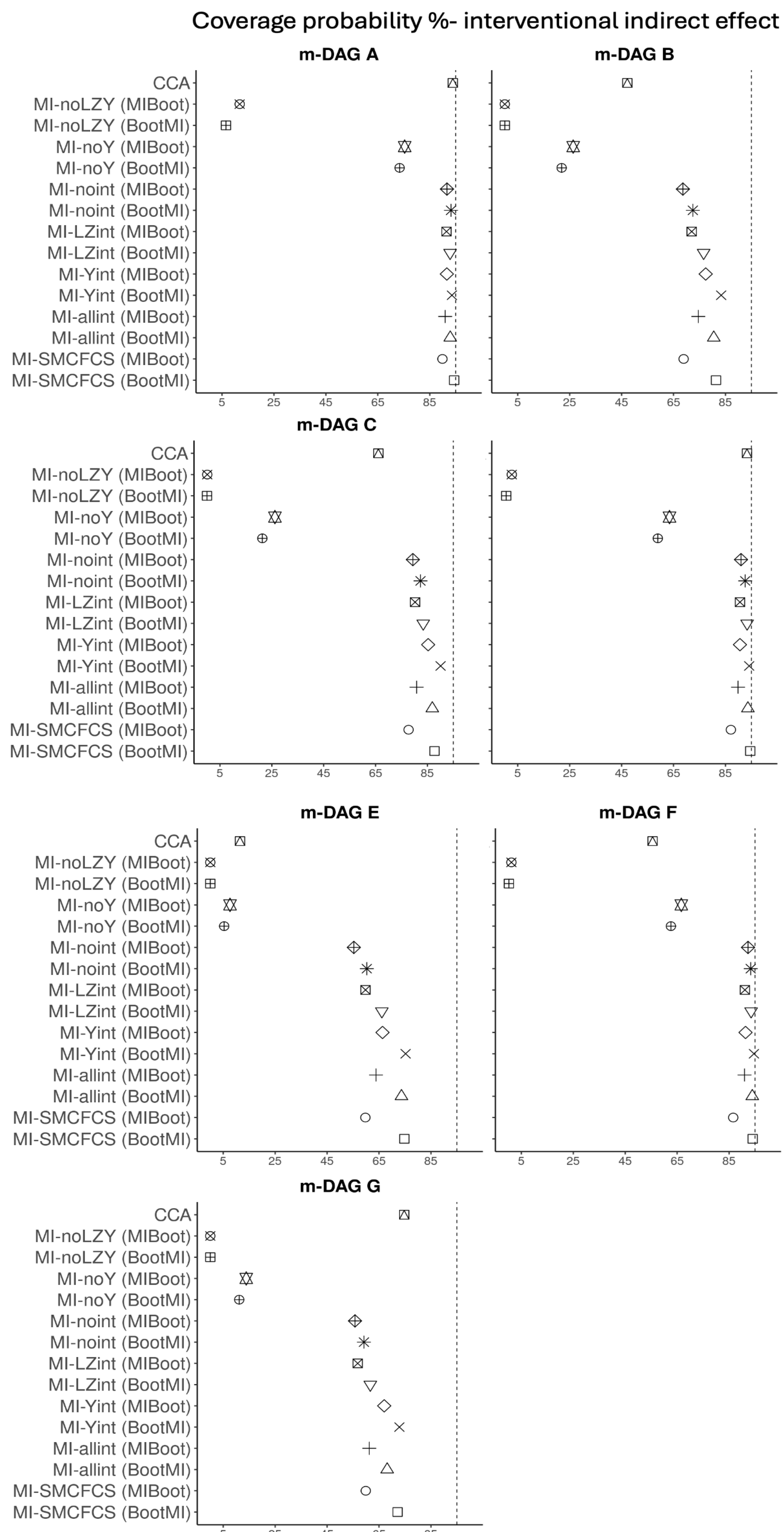

Note: See Table 2 for description of each of the approaches used for handling missing data. Due to perfect prediction, the MI-SMCFCS (BootMI) approach failed in 13 out of the 2,000 simulated datasets under m-DAG B and 7 out of 2,000 under E. The Monte Carlo standard errors ranged from 0% to 1.12% (eTable 4).

eFigure 4. Coverage probability (%) in estimates of the interventional direct effect across the m-DAGs using Monte Carlo simulation-based g-computation as substantive mediation analysis method and different approaches to handle missing data.

Coverage probability %- interventional direct effect

Note: See Table 2 for description of each of the approaches used for handling missing data. Due to perfect prediction, the MI-SMCFCS (BootMI) approach failed in 13 out of the 2,000 simulated datasets under m-DAG B and 7 out of 2,000 under E. The Monte Carlo standard errors ranged from 0.46% to 1.11% (eTable 4).

eFigure 5. Estimated effect of a hypothetical intervention on CMD in young adulthood on reducing CMD in mid-adulthood in individuals with adolescent CMD (interventional indirect effect) and the remaining effect of adolescent CMD (interventional direct effect), using a Monte Carlo simulation-based g-computation approach with different approaches to handling missing data within the VAHCS case study

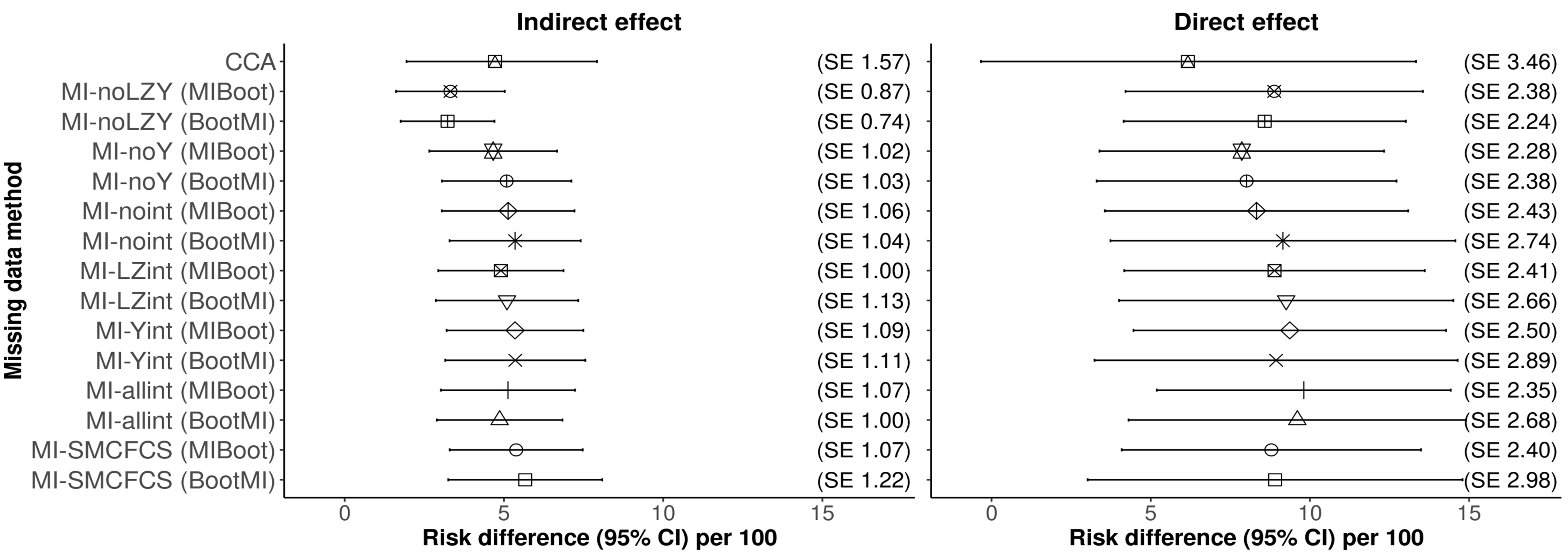

In all analyses, the MC g-computation was implemented using 1,000 Monte Carlo draws. The BootMI approaches were conducted using 200 bootstrap samples and two imputed datasets, and the MIBoot approaches using 50 imputed datasets and 200 bootstrap samples. Following the original investigation,(23) sex, cannabis use, anti-social behavior, incomplete high school, parental completion of high-school, parental divorce, and socioeconomic disadvantage, all measured across adolescent waves (waves 2-6) were included as baseline confounders in the analysis. Parenthood by age 2 (wave 8) was included in the analysis as intermediate confounder. We included parent smoking and drinking by wave 7 and participants drinking and smoking at each of waves 2 to 6 as auxiliary variables in the multiple imputation approaches. There were missing data in some of the auxiliary variables (ranging from 5% to 21%) and additional baseline confounders (ranging from 0.6% to 7%). These variables were also multiply imputed in the all multiple imputation approaches.